\documentclass[aip,reprint]{revtex4-1}
\usepackage{color} 
\usepackage{setspace}
\usepackage{amssymb,amsmath} 
\usepackage{graphicx}
\usepackage[caption=false]{subfig}
\usepackage{ulem}
\usepackage{wrapfig}
\usepackage{enumerate}

\begin{document}

\title{The fundamental role of quantized vibrations in coherent light harvesting by cryptophyte algae}
\author{Avinash Kolli}
\affiliation{Department of Physics and Astronomy, University College London, Gower Street, London, WC1E 6BT United Kingdom.}
\author{Edward J. O'Reilly}
\affiliation{Department of Physics and Astronomy, University College London, Gower Street, London, WC1E 6BT United Kingdom.}
\author{Gregory D. Scholes}
\affiliation{Department of Chemistry, Institute for Optical Sciences and Centre for Quantum Information and Quantum Control, University of Toronto, 80 St. George Street, Toronto, Ontario, M5S 3H6 Canada.}
\author{Alexandra Olaya-Castro}
\email{a.olaya@ucl.ac.uk}
\affiliation{Department of Physics and Astronomy, University College London, Gower Street, London, WC1E 6BT United Kingdom.}


\begin{abstract}

The influence of fast vibrations on energy transfer and conversion in natural molecular aggregates is an issue of central interest. This article shows the important role of  high-energy quantized vibrations and their non-equilibrium dynamics for energy transfer in photosynthetic systems with highly localized excitonic states. We consider the cryptophyte antennae protein phycoerythrin 545 and show that coupling to quantized vibrations which are quasi-resonant with excitonic transitions is fundamental for biological function as it generates non-cascaded transport with rapid and wider spatial distribution of excitation energy. Our work also indicates that the non-equilibrium dynamics of such vibrations can manifest itself in ultrafast beating of both excitonic populations and coherences at room temperature, with time scales in agreement with those reported in experiments. Moreover, we show that mechanisms supporting coherent excitonic dynamics assist coupling to selected modes that channel energy to preferential sites in the complex. We therefore argue that, in the presence of strong coupling between electronic excitations and quantized vibrations, a concrete and important advantage of quantum coherent dynamics is precisely to tune resonances that promote fast and effective energy distribution.  
\end{abstract}

\maketitle

\definecolor{purple}{RGB}{102,0,153}

\section{Introduction}
The functionality of natural multichromophoric complexes depends fully on the strength and structure of interactions between electronic states and vibrational degrees of freedom associated either to intramolecular  vibrations or to intermolecular interactions with the protein and the surrounding solvent environment \cite{renger01, amerongen00, roden09, hoda12}. In these systems, the phonon environment influencing electronic excitation dynamics can be broadly divided into two distinct components \cite{renger01}. Firstly, a continuous distribution of modes with a characteristic cutoff frequency below or comparable to the thermal energy scale, $k_{B} T$. These modes, arising from the solvent and low-energy protein vibrations, induce thermal fluctuations of onsite energies which in turn drive transitions between energetically close excitonic states. Secondly, spectroscopy studies \cite{womick09, west11, ratsep08, wendling00, raja93, jankoviak11, eisenmayer12, womick11, richards12} also reveal the active participation of  specific  vibrational modes during excitation dynamics, often described as peaks in the spectral density characterising exciton-phonon interactions.  Vibrations with lower frequencies may arise from inter-molecular coupling  between pigments and the local protein environment \cite{eisenmayer12} while  peaks at high energies  originate mostly from intramolecular vibrations that give rise to vibronic transitions \cite{davidov62, yang06}. We denote these high-energy vibrational modes as \textit{quantized vibrations} to indicate that their quantum mechanical features can be significant even at ambient temperatures \cite{renger01}. In contrast,  vibrations with frequencies below $k_{B} T$ will be thermally activated.  Indeed, the influence of quantized vibrations in energy transport in a variety of natural molecular systems is currently of central interest \cite{womick11, richards12, turner12, ritschel12, hay12}. 

In  this work we discuss the important role of quantized vibrations and their non-equilibrium dynamics for electronic excitations in the cryptophyte antennae protein phycoerythrin 545 (PE545) \cite{doust04, wilk99}, where exciton coherence beating has been observed \cite{collini10}.  Chromophores in cryptophyte light-harvesting antenna proteins have large energy gaps and are separated by an average distance of $20~\textrm{\AA}$ \cite{doust04}, which is double the average pigment distance in proteins of plants \cite{amerongen00}, hence the majority of electronic couplings in PE545 are small ($<$$30\, \textrm{cm}^{-1}$) \cite{doust04}.  How can this antenna protein function efficiently with such structure?  We show that a fundamental mechanism for the biological function of this complex and for supporting coherence oscillations at long-times is the coupling between excitons and vibrations quasi-resonant with excitonic energy differences. The importance of these interactions in similar light-harvesting systems has recently been highlighted by ultrafast spectroscopy experiments \cite{womick11, richards12, turner12}.

Vibrations reported in spectroscopy of PE545 have high-frequencies \cite{doust04, novoderezhkin10} with the largest couplings for modes between $514\, \rm{cm}^{-1}$ and $1450\,\rm{cm}^{-1}$. Given that these frequencies match closely energy gaps between exciton states, transport in PE545 is fundamentally activated by these vibrations giving rise to the transfer pathways and transfer times observed in experiments \cite{novoderezhkin10}. This result can be anticipated using F\"orster theory \cite{forster} but is derived here using a non-Markovian polaron-representation master equation for multichromophoric systems \cite{kolli11, jang11}. This more general framework allows us to investigate the evolution of excitonic coherences and to show that non-equilibrium dynamics associated to such vibrations can contribute to the long-time oscillations of the coherences probed in recent ultrafast experiments \cite{collini10, turner12}. Based on this analysis, we are able to elucidate two key features of excitation transfer assisted by vibrations in cryptophytes. First, excitation dynamics follows a non-sequential order bridging large energy gaps and distances and therefore generating faster and wider spatial distribution of excitation energy. Second, vibrations that modulate the evolution of excitonic coherences also promote energy transfer to preferential sites in the complex. Both of these process support biological function of the antennae by assuring that energy transfer to other photosynthetic complexes happens in the relevant biological time scale. On the basis of these results we argue that in complexes with highly localized excitons a fundamental role for quantum coherent contributions to dynamics is precisely to ensure that resonances with selected vibrations are tuned to guarantee faster and effective distribution of excitation energy. The interactions between excitons and quantized vibrations in close resonane with exciton gaps are therefore an important design principle for optimal light-harvesting antennae\cite{scholes11}.
\begin{figure*}[]
\centering 
\subfloat[{Model of PE545 showing the four protein subunits as colored ribbons}]{
\includegraphics[width=75mm]{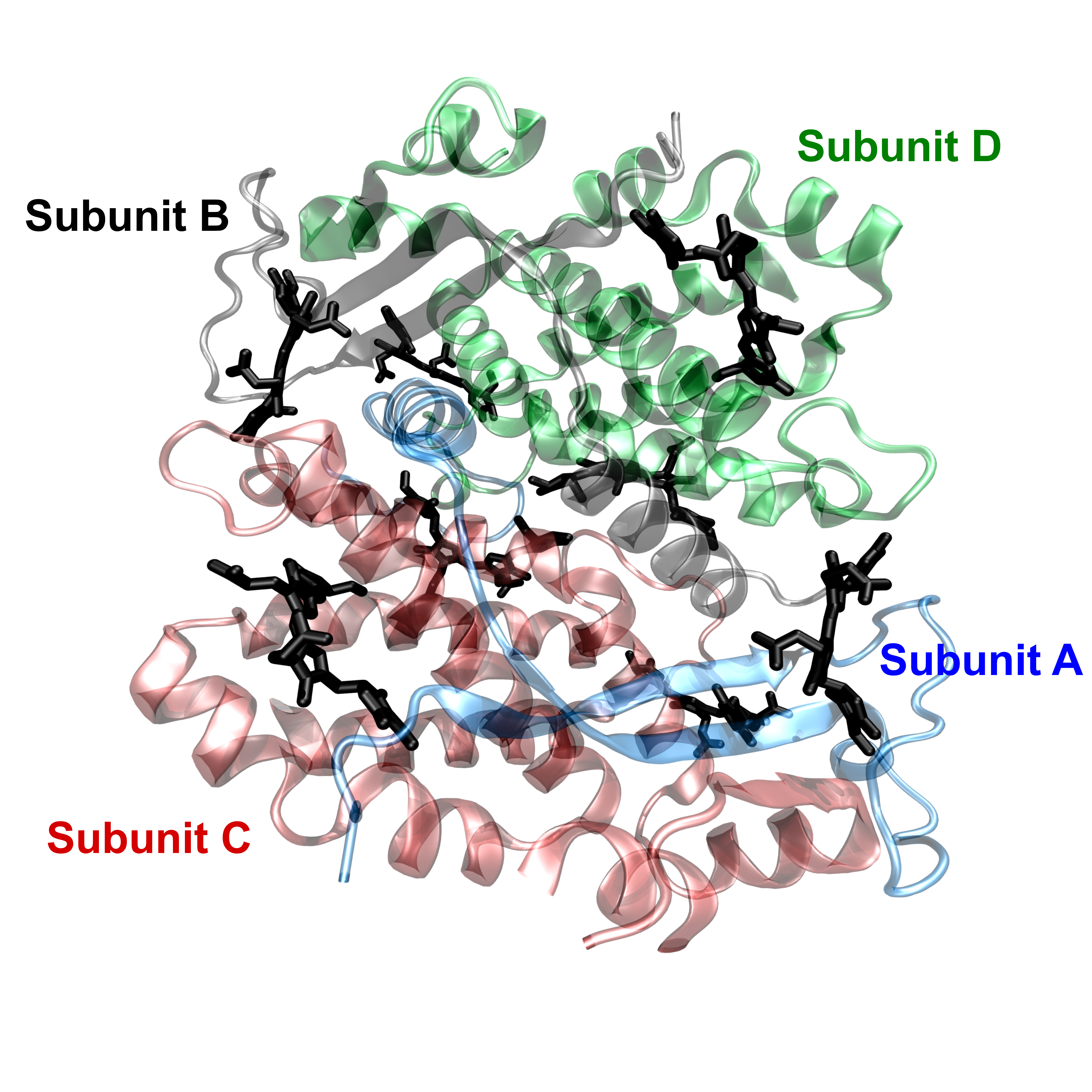}
\label{fig:subunits}
}
\subfloat[Structure of the isolated pigment molecules in PE545]{
\includegraphics[width=75mm]{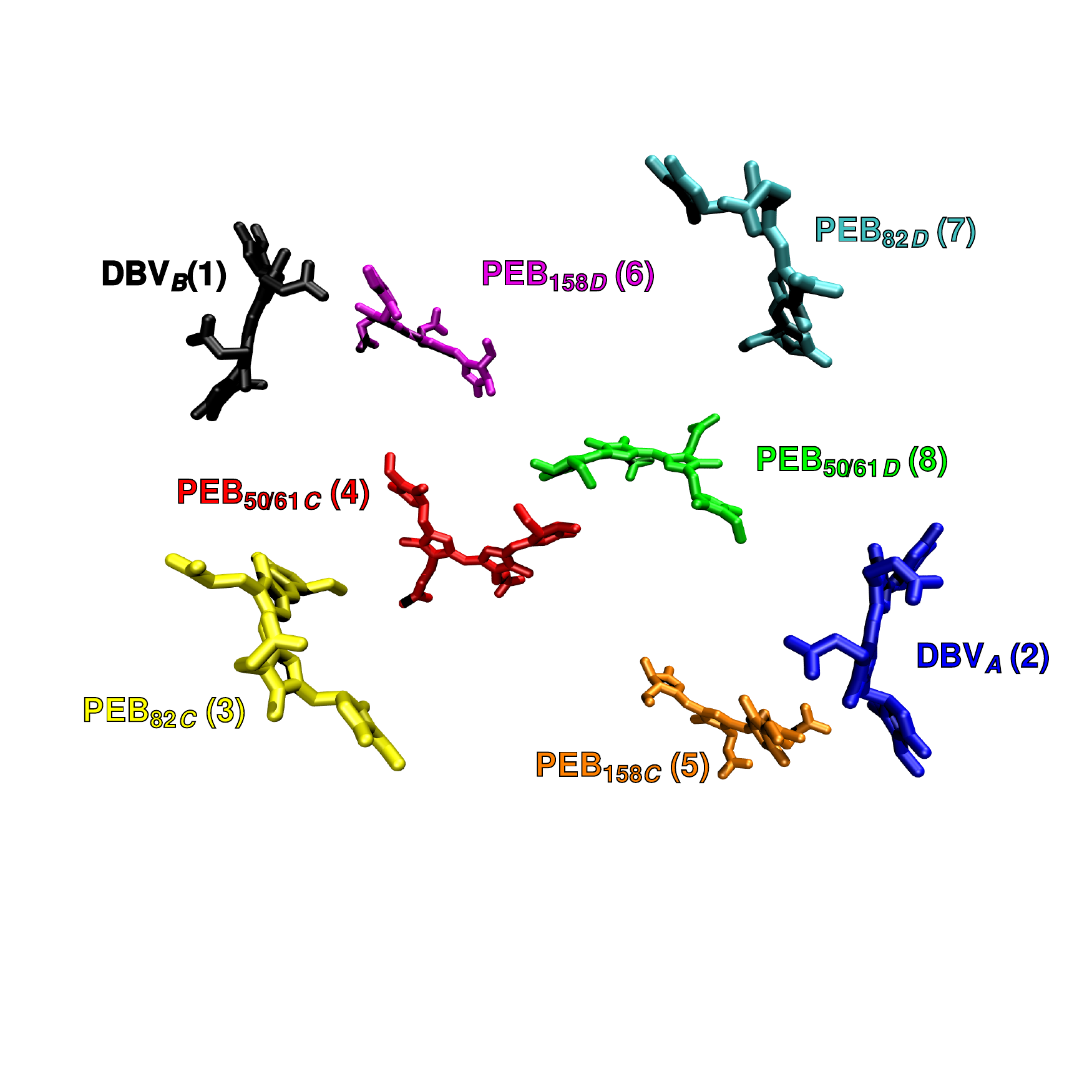}
\label{fig:chromophores}
} \\
\subfloat[Table showing bare excitation energies for each pigment and the protein subunit to which it is bounded. These values have been taken from Ref. \cite{novoderezhkin10} and they correspond to a model that best fit experimental results.]{
\includegraphics[width=150mm]{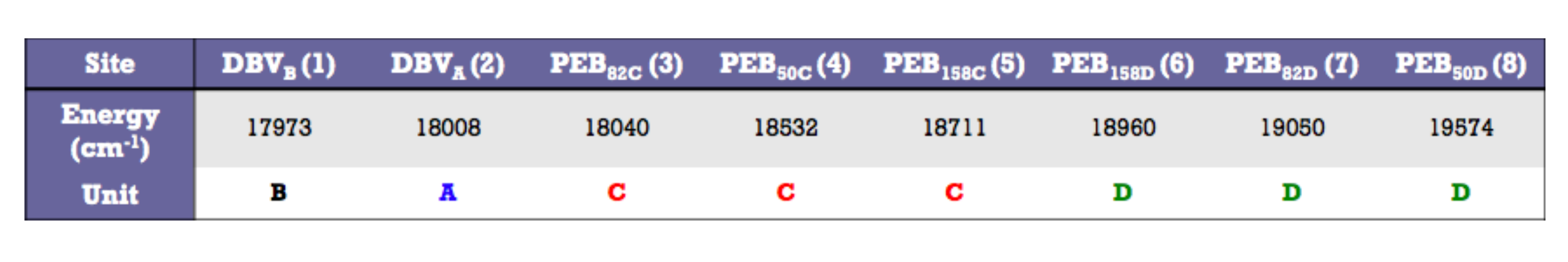}
\label{fig:table1}
} \\
\caption{Model of an isolated PE545 from cryptomonad
\textit{Rhodomonas} CS24 based on the X-Ray structure at 0.97\AA~resolution (Protein Data Bank ID code 1XG0)\cite{wilk99, doust04} }
\label{fig:structures}
\end{figure*}

\section{Marine Cryptophyte Antennae Proteins}
Cryptophyte photosynthesis relies on phycobiliproteins such as PE545 and
Phycocyanin 645 (PC645) to ensure photon absorption in the visible part of the spectrum, with an optimal performance in the blue ($545$ nm) and green ($645$ nm) regions \cite{bernstein89, doust04}. Excitation energy absorbed by cryptophyte antennae proteins is transferred towards photosystems II and II (PSI and PSII) \cite{wit06}. Crystallography \cite{wilk99,doust04} reveals that PE545 contains eight light-absorbing bilin molecules covalently bound to a four subunit protein scaffold labelled A, B, C and D (see Figure \ref{fig:subunits}). The lowest-energy chromophores are 15,16-dihydrobiliverdin chromophores and are linked to separate  protein subunits A and B ($\rm{DBV_{A}}$ and $\rm{DBV_{B}}$). The remaining six chromophores are phycoerythrobilin (PEB) bilins, two of which are doubly covalently bound to the protein environment while the remaining four are singly bound. The PEB bilins are arranged such that pigments $\rm{PEB_{50/61C}}$, $\rm{PEB_{82C}}$ and $\rm{PEB_{158C}}$ occupy the subunit C, while chromophores $\rm{PEB_{50/61D}}$, $\rm{PEB_{82D}}$ and $\rm{PEB_{158D}}$ are attached to  subunit D. One particular feature of this complex is that it contains a deep, water-filled slot exposing the pigments to a large contact with water. The combined pigment-protein structure is shown in Figure \ref{fig:subunits}, while the relative positions of the isolated pigments is illustrated in Figure \ref{fig:chromophores}. 

\subsection{The Electronic System}
Quantum chemical calculations \cite{doust04, curutchet11} as well as spectroscopy \cite{doust04, novoderezhkin10} have allowed the refinement of a parametrized Hamiltonian for the electronic structure of PE545 given by ($\hbar=1$):
\begin{eqnarray}
H_{el} &=& \sum_{m} \epsilon_{m} \sigma_{m}^{+} \sigma_{m}^{-} + \sum_{\langle m, n \rangle} V_{mn}   (\sigma_{m}^{+} \sigma_{n}^{-} + \sigma_{n}^{+} \sigma_{m}^{-}),
\label{eq:hel}
\end{eqnarray}
with $\sigma_{m}^{+}$ being the operator that creates an excitation on site $m$ i.e. $\sigma_{m}^{+}|0\rangle=|m\rangle$.  A salient characteristic of this system is that site energy differences $|\epsilon_{m}-\epsilon_{n} |$ are very large in comparison to inter-site electronic interaction $V_{mn}$, for instance, the largest coupling is $V_{48}=92\,\rm{cm}^{-1}$ between the central pigments $\rm{PEB_{50C}}$  and $\rm{PEB_{50D}}$ while their energy difference is about $1040~\rm{cm}^{-1}$ (remaining values of $V_{mn}$ can be found in the supplemental information). As a result, each excitonic eigenstate, which we denote $|e_m\rangle$, is highly localized to a particular site $m$. The table in Figure \ref{fig:table1} presents the site energies of the eight chromophores in ascending order as well as the subunits to which each of them is bound. Notice the interesting relationship between the spatial arrangement of the pigments and their energies: the three highest energy chromophores are located in subunit D, the next three highest energy chromophores lie in subunit C and finally the two lowest chromophores reside in subunits B and A. Therefore, the hierarchy of pigment energies imposes also a hierarchy of energies among protein subunits. In section \ref{sec:role} we illustrate how one of the functions of quantized vibrations is precisely to support efficient energy transfer  between protein subunits. 

\subsection{System-Environment interaction}
\label{sec:env}
Optimal energy transfer in light harvesting systems depends crucially on the transitions and correlations that can arise from interaction between the electronic system and their environment. The later is usually described as a bath of harmonic modes of frequencies $\omega_{\mathbf{k}}$, with boson creation operators $b_{\mathbf{k}}^{\dagger}$ and Hamiltonian  $H_{B}=\sum_{\mathbf{k}} \omega_{\mathbf{k}} b_{\mathbf{k}}^{\dagger} b_{\mathbf{k}} $. The linear interaction of this harmonic bath and the electronic system is given by
\begin{eqnarray}
H_{I} &=& \sum_{m} \sigma_{m}^{+} \sigma_{m}^{-} \sum_{\mathbf{k}} (g_{\mathbf{k},m} b_{\mathbf{k}}^{\dagger} + g_{\mathbf{k},m}^{*} b_{\mathbf{k}}),
\end{eqnarray}
where $g_{\mathbf{k},m}$ is the coupling of site $m$ to mode $\mathbf{k}$. The strength and structure of these interactions is  characterized by the spectral density, $J(\omega)=\sum_{\mathbf{k}} |g_{\mathbf{k}}|^{2} \delta(\omega-\omega_{\mathbf{k}})$ which has two contributions: coupling to low-frequency continuous modes denoted as $ J_{CM}(\omega)$ and coupling to quantized vibrations denoted  $J_{QM}(\omega)$  such that $J(\omega) = J_{CM}(\omega) + J_{QM}(\omega)$.  

Guided by linear spectra studies on PE545 \cite{doust04, novoderezhkin10}, in this work we consider that $ J_{CM}(\omega) $ is given by a super-Ohmic spectral density of the form
\begin{eqnarray}
J_{CM}(\omega) &=& \frac{s_{1} \omega^3}{2 \Omega_{1}^2} e^{-(\omega/\Omega_{1})^{\frac{1}{4}}} + \frac{s_{1} \omega^3}{2 \Omega_{1}^2} e^{-(\omega/\Omega_{1})^{\frac{1}{4}}}.
\end{eqnarray}
\noindent Here the parameters (in units of $\rm{cm}^{-1}$) for the continuous modes are $s_{1}=3.46 \times 10^{-4}$, $s_{2}=2.02 \times 10^{-4}$, $\Omega_{1}=1.45 \times 10^{-3}$ and $\Omega_{2}=4.34 \times 10^{-3}$. We have chosen a super-Ohmic form for the continuous component of the spectral density taking into account that the frequency dependence of this function has been shown to appropriately reproduce vibrational sidebands \cite{renger02}. Moreover, the theoretical framework that we will use for describing dynamics is more accurate in this case \cite{kolli11, jang11}. 

\begin{figure*}[]
\centering
\subfloat[Spectral density and excitonic transitions]{
\includegraphics[width=95mm]{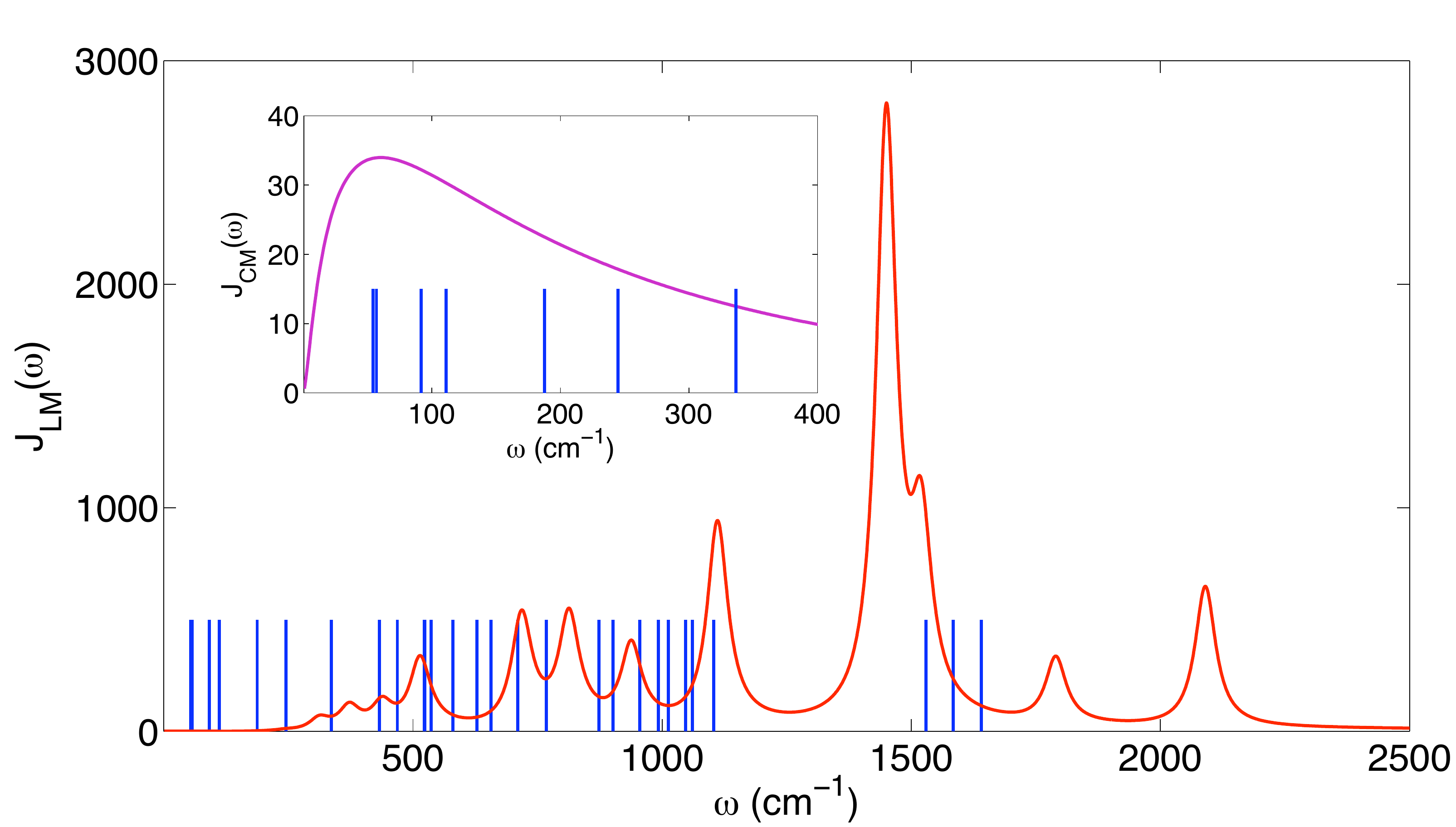}
\label{fig:spectraldensities}
}
\subfloat[Schematic of prototype pairs in PE545 ]{
\includegraphics[width=60mm]{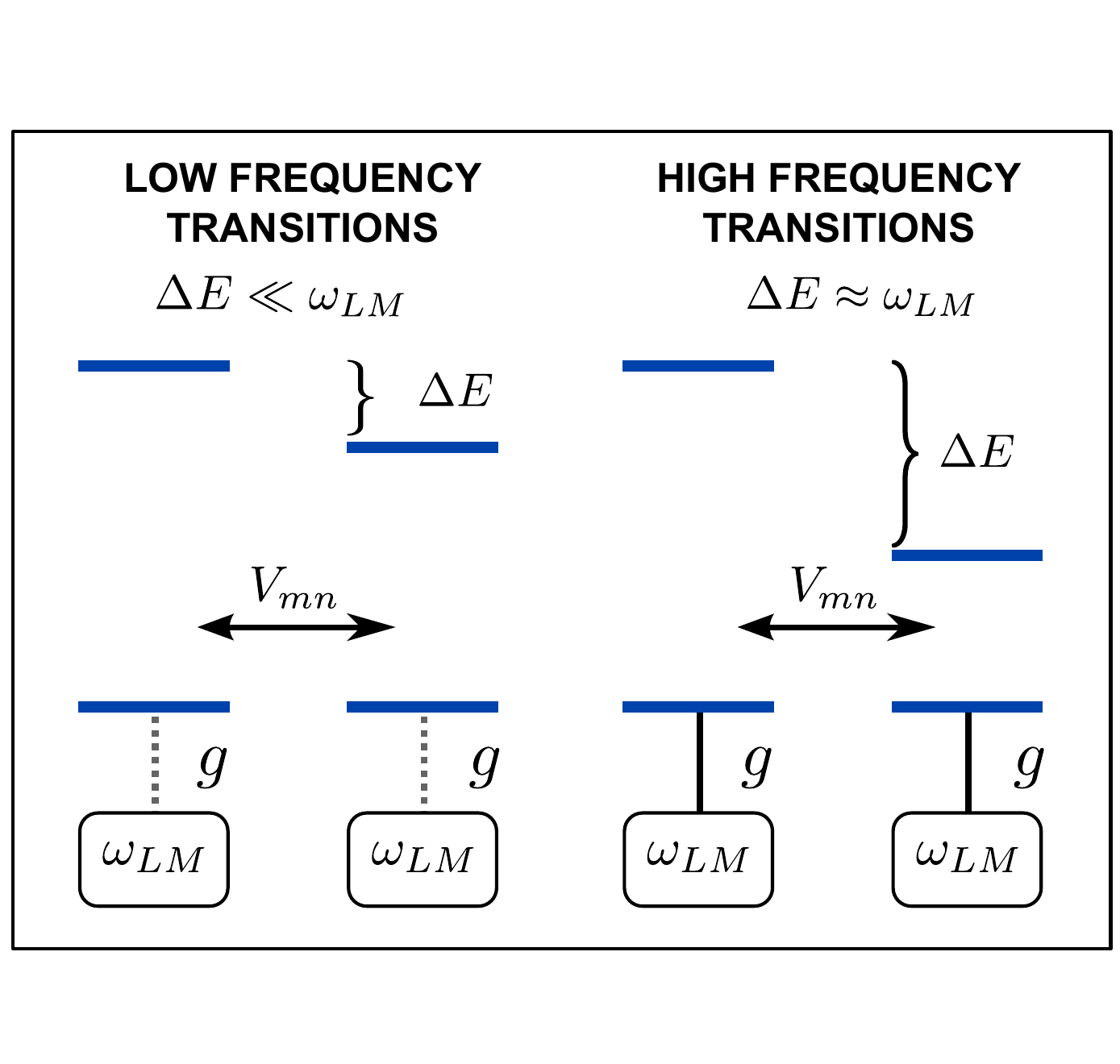}
\label{fig:highenergymodes}
}
\caption{(a) Main panel shows the contribution of quantized vibrations $J_{QM}(\omega)$, while the inset depicts the continuous spectral density $J_{CM}(\omega)$. Bars indicate excitonic transitions. (b) Some pigment pairs have energy detunings $\Delta E$ much smaller than any of the $\omega_{QM}$, for instance the pair $\rm{DVB_{A}}$ and $\rm{DVB_{B}}$ (sites 1 and 2) for which $\Delta E=35~\rm{cm}^{-1}$, thus quantized vibrations have a small effect on this excitonic transition. In contrast, many pairs formed of pigments located on different protein subunits have $\Delta E$ in close resonance with a vibrational mode, example of this is the central dimmer $\rm{PEB_{50C}}$ and $\rm{PEB_{50D}}$ (sites 4 and 8) whose energy difference is $1040 \rm{cm}^{-1}$. In this later case, electronic interaction $V_{mn} $ is the smallest energy scale.}
\end{figure*}

Spectroscopy \cite{doust04} also reveals the influence of 14 vibrational components in the excitation dynamics on PE545, whose frequencies $\omega_{QM}$ and Huang-Rhys factors $s_{QM}$ are shown in Table \ref{tab:pe545highmodes}. Notice that the majority of the vibrations satisfy the condition $\hbar \omega_{QM}\geq k_BT\sim 200 {\rm cm}^{-1}$ falling in the regime of high-energy. The largest couplings (given by $g=\omega_{QM}\sqrt{s_{QM}}$) are for modes $\omega_{QM}=1111~{\rm cm}^{-1}$ and $\omega_{QM}=1450~{\rm cm}^{-1}$.  In this work we include  the effect of these modes via a spectral density exhibiting sharp components. When combined with the non-Markovian polaron treatment (see \ref{sec:polaron}), this provides a starting point to investigate the effects of non-equilibrium dynamics associated to these vibrations. In practice, these modes could be broadened by interactions with the surrounding environment.  Therefore, in this work we assume broadened vibrational modes, each with a Lorentzian line shape of broadening $b$ \cite{garg85}. The contribution of each of these modes to the spectral density is given by:
\begin{equation}
J_{QM}(\omega) = \frac{2 s_{QM} \omega_{QM}}{\pi} \frac{\omega^{3} b}{(\omega^{2}-\omega_{QM}^{2})^{2} + b^{2}\omega^{2}} \,. 
\end{equation}
A first insight into the relevance of the different components of $J (\omega)$ in excitation dynamics can be gained by understanding the relative position of excitonic transitions to the spectral density. Figure \ref{fig:spectraldensities} shows the forms of $J_{CM}(\omega) $ and $J_{QM}(\omega)$ as well as superimposed  bars indicating the excitonic energy differences of the the electronic Hamiltonian (Eq. \ref{eq:hel})).  One striking feature in this figure is that the energies of the vibrations closely match a large number of the excitonic transitions. For example, the localized mode at $\omega_{QM}=1111~\textrm{cm}^{-1}$ lies close to resonance with the transition between eigenstates $|e_{8}\rangle$ and $|e_{4}\rangle$ and also to the transition between eigenstates $|e_{7}\rangle$ and $|e_{2}\rangle$. To understand the physical significance of these resonances, consider a scenario where a single localized mode is exactly on resonance with an excitonic transition. In this case, the electronic system is able to emit a phonon into the mode, preserving energy during the transition. Therefore, quasi-resonant vibrations can activate otherwise energetically unfavorable excitonic transitions, bridging large energy gaps and, in this case, also large distances. 

\begin{table}[]
\begin{center}
\begin{tabular}{| l || ccccc |}
\hline
$s_{QM}$ & 0.0013 & 0.0072 & 0.045 & 0.0578 & 0.045 \\ 
$\omega_{QM}$ & 207 & 244 & 312 & 372 & 438 \\
\hline
$s_{QM}$ & 0.0924 & 0.0761 & 0.0578 & 0.0313 & 0.0578  \\ 
$\omega_{QM}$ & 514 & 718 & 813 & 938 & 1111 \\
\hline
$s_{QM}$ & 0.1013 & 0.0265 & 0.0072 & 0.0113 & \\
$\omega_{QM}$ & 1450 & 1520 & 1790 & 2090 &\\
\hline
\end{tabular}
\caption{Vibrational parameters for PE545 taken from Ref. \cite{doust04}. Frequencies in units of $\textrm{cm}^{-1}$.} 
\label{tab:pe545highmodes}
\end{center}
\end{table}

\textbf{Environmental Correlations:} 
The mechanisms supporting beating of excitonic coherences in  light-harvesting systems are still under  and they can indeed be system-dependent. In the case of PE545, it has been argued that non-Markovian dynamics arising from slow relaxing modes or solvent-induced correlated fluctuations of on-site energies can help supporting long-lasting oscillations of excitonic coherences\cite{collini10}. So far however none of these hypothesis has been experimentally probed. Indeed, atomistic descriptions have challenged the idea of spatially correlated fluctuations in other molecular complexes \cite{olbrich11}. However, solvent-induced correlated fluctuations can be expected to play a more significant role in systems where chromophores are in close proximity to water. An example is hydrated DNA where hole transport is affected by correlated fluctuations that arise from the nearby solvent \cite{kubar09}. Interestingly, the structure of PE545  is such that a deep pocket of bound water occupies space between the monomers, thus allowing close contact of the pigments with the water molecules. Atomistic calculations to understand whether interaction with near-by solvent can induce correlated fluctuations in PE545 are under development \cite{benedeta}. Therefore, in this work we compute linear spectra and dynamics in PE545 with and without solvent-induced correlated fluctuations. 
 
Spatially correlated fluctuations are introduced via position dependent coupling to common propagating modes, i.e. $g_{\mathbf{k},m} = g_{\mathbf{k}} e^{i \mathbf{k}.\mathbf{r}_{m}}$ where $\mathbf{r}_{m}$ denotes the position of the $m$'th chromophore \cite{fassioli10, nalbach10}. Generally, this behaviour is limited to wavelengths on which the environment appears homogeneous, restricting the modes to the low energy sector. Therefore, we assume that only the continuous component of the spectral density induces correlated fluctuations. For an isotropic three-dimensional bath,  the spectral density function of fluctuations between sites $m$ and $n$ can be written as $J^{mn}(\omega) = J_{CM}^{mn}(\omega) + J_{QM}^{mn}(\omega)$, where 
$J_{CM}^{mn}(\omega) = J_{CM}(\omega) ~\textrm{sinc}(\omega r_{mn}/v_{ph}) $ and $J_{QM}^{mn}(\omega) =  J_{QM}(\omega) \delta_{mn}$. Here $r_{mn}$ corresponds to the distance between the $m$'th and $n$'th chromophores, while $v_{ph} =1000~\textrm{ms}^{-1}$ is comparable to the speed of sound in water. If instead $v_{ph}$ is set to zero, we recover the case of independent harmonic baths. For a detailed justification of this model of solvent-induced correlations see Nalbach et al. \cite{nalbach10}.

\section{Dynamics and spectra in the polaron-representation}
\label{sec:polaron}
Exact treatment of the exciton-phonon interactions across the whole range of energy and couplings is not a straightforward task \cite{silbey84, jang08, ishizaki09, nazir09, olaya11, kolli11, jang11, nalbach11}. Importantly, in the regime of strong system-environment coupling, F\"orster theory  \cite{forster} accounts well for the influence of high-energy vibrations in transfer rates. However, F\"oster theory fails to capture the non-equilibrium dynamics of such vibrations due to interactions with the electronic components. A  formal treatment of quantized vibrations consider them as part of what is denoted as \textit{system}\cite{renger01}  such that coherent quantum interactions between electronic and vibrational modes are accounted for. We illustrate such an approach with an example in section VC but the full quantum treatment of PE545 with 14 vibrational modes per pigment is beyond the scope of this paper.  Instead, here we use a non-Markovian many-site polaron master equation \cite{kolli11, jang11}, which allows consideration of  non-equilibrium dynamics of high-energy quantized vibrations, that is, the fact that these modes take some time to thermalize. As we will see here, these non-equilibrium dynamics induce modulation of excitonic population and coherences. The treatment is particularly accurate for systems where excitonic states are highly localized and hence suitable for describing dynamics and spectroscopy of PE545. Full details of the polaron master equation for multichromophoric systems are given in Refs. \cite{kolli11, jang11}. Notice that this formalism allows reconstruction not only of the site-population dynamics as originally proposed \cite{jang08, jang11} but also of the evolution of excitonic coherences \cite{kolli11}. For self-consistency of the paper we summarize the framework here. 

The polaron formalism describes a dynamics in a framework of displaced oscillators where the original Hamiltonian $H=H_{el}+H_{B}+H_{I}$ is transformed via a small polaron-transformation i.e. $\tilde{H}=e^{S}He^{-S}$ with $S=\sum_{m} \sigma_{m}^{+} \sigma_{m}^{-} \sum_{\mathbf{k}}  (h_{\mathbf{k},m} b_{\mathbf{k}}^{\dagger} - h_{\mathbf{k},m}^{*} b_{\mathbf{k}})$ and $h_{\mathbf{k},m} = g_{\mathbf{k},m}/\omega_{\mathbf{k}}$. In this shifted-oscillators framework,  the effective  electronic system is described by
\begin{eqnarray}
\tilde{H}_{el} &=& \sum_{m} \tilde{\epsilon}_{m} \sigma_{m}^{+} \sigma_{m}^{-} + \sum_{\langle m, n \rangle} V_{mn}  \beta_{mn} (\sigma_{m}^{+} \sigma_{n}^{-} + \sigma_{n}^{+} \sigma_{m}^{-}).\nonumber\\
\label{eq:heldressed}
\end{eqnarray}
Here $\tilde{\epsilon}_{m}$ and $V_{mn}  \beta_{mn}$ are shifted onsite energies and renormalized electronic interactions respectively. The eigenstates of $\tilde{H}_{el}$ are labelled $|\alpha_m\rangle$ and are denoted as renormalized exciton states from now on.  Interaction with the bath of shifted oscillators induce fluctuations of electronic couplings given by 
\begin{eqnarray}
\tilde{H}_{I}&=&\sum_{\langle m, n \rangle} V_{mn}  (\tilde{B}_{mn} \sigma_{m}^{+} \sigma_{n}^{-} + \tilde{B}_{mn}^{\dagger} \sigma_{n}^{+} \sigma_{m}^{-}).
\label{eq:polaronH}
\end{eqnarray}
Bath-induced renormalisation factors are defined as $\beta_{mn} = \langle B_{mn} \rangle$ while displaced bath operators become $\tilde{B}_{mn} = B_{mn} - \beta_{mn}$ with $ B_{mn} = e^{\sum_{\mathbf{k}}(\delta h_{\mathbf{k},m n} b_{\mathbf{k}}^{\dagger} - \delta h_{\mathbf{k},m n}^{*} b_{\mathbf{k}})}$. When a common bath is assumed, the factor $\delta h_{\mathbf{k},m n} = h_{\mathbf{k},m} -h_{\mathbf{k},n}$ depends on the difference in electron-phonon couplings of sites $m$ and $n$. For a harmonic oscillator bath in thermal equilibrium, the renormalisation factors evaluate to $\beta_{mn} = e^{-\frac{1}{2}\sum_{\mathbf{k}} \coth (\beta\omega_{\mathbf{k}}/2) |\delta\alpha_{\mathbf{k},mn}|^{2}}$.

Perturbation theory with respect to $\tilde {H}_{I}$, leads to a non-Markovian polaron master equation 
\begin{equation}\label{eq:TCL}
\frac{d}{dt} \tilde{\rho}(t) = \mathcal{R}(t)\tilde{\rho}(t) + \mathcal{I}(t) \tilde{\rho}(0),
\end{equation}
The homogeneous term $\mathcal{R}(t)\tilde{\rho}(t)$ describes the relaxation of the electronic degrees of freedom due to the interaction with a thermalised environment. The inhomogeneous term $\mathcal{I}(t) \tilde{\rho}(0)$ is only non-zero when the initial environmental state differs from the thermal state in the polaron frame. Therefore, this term allows us to capture the non-equilibrium dynamics of all vibrational components and in particular the fact that high-energy quantized vibrations are taken out of thermal equilibrium due to interaction with the electronic component \cite{kolli11, jang11}. Explicit expressions for the homogeneous and inhomogeneous terms are presented in the supplemental information. 
 
\textbf{Original frame observables}
The master equation presented in Eq. (\ref{eq:TCL}) describes the evolution of system observables within the polaron frame ($\tilde{\rho}(t)$). However, we are ultimately interested in the dynamics as seen in the  original untransformed `lab' frame ($\rho(t)$) since this will be the appropriate operator to compute expected values of physical observables. A general prescription for the transformation of system observables from the polaron to the lab frame was presented in Kolli {et al.} \cite{kolli11}. These transformations can be summarised as follows: site populations $\langle \sigma_{m}^{+} \sigma_{m}^{-} \rangle$ remain unaffected upon transformation to the lab frame as the operator $\sigma_{m}^{+} \sigma_{m}^{-}$ commutes with the polaron transformation $S$.  Expected values of off-diagonal operators in the site basis are given by an elaborated transformation, which up to first order in $\tilde{H}_I$ reads: 
\begin{eqnarray}
\langle \sigma_{m}^{+} \sigma_{n}^{-} \rangle &=& \textrm{tr}\{ \sigma_{m}^{+} \sigma_{n}^{-} \rho(0) \} + \beta_{mn} \textrm{tr}\{ \sigma_{m}^{+} \sigma_{n}^{-} [\tilde{\rho}(t)- \tilde{\rho}(0)]\}  \nonumber\\
&&+ \textrm{tr} \{ \sigma_{m}^{+} \sigma_{n}^{-} S_{1}(t) \tilde{\rho}(t) \} 
+ \textrm{tr}\{\sigma_{m}^{+} \sigma_{n}^{-} T_{1}(t) \tilde{\rho}(0) \}. \nonumber
\end{eqnarray}
Expression for super-operators $S_{1}(t)\tilde{\rho}(t)$ and $T_{1}(t)\tilde{\rho}(0)$ are presented in the supplemental information. 

\textbf{Linear Spectra:} 
Here we provide a derivation of the linear spectra based on the polaron-representation of the dynamics. Our derivation of the spectra appeals to the Markov and secular approximations \textit {in the polaron frame}, formulation which may still capture some non-Markovian effects in the untransformed frame (see details in Kolli \textit{et al.} \cite{kolli11}). The absorption spectra of a multichromophoric system is defined in terms of Fourier transform of the dipole-dipole correlation function \cite{may04}:
\begin{eqnarray}
A(\omega) = Re \int_{0}^{\infty} dt~e^{i\omega t}~C_{d-d}(t) 
\end{eqnarray}
\noindent The dipole-dipole correlation function is defined as the two-time correlation function of the transition dipole operator
\begin{equation}
C_{d-d}(t) = \textrm{Tr}_{S+B} \{ \mu(t) \mu(0) \rho_{B} |0\rangle\langle 0| \}
\end{equation}
\noindent where $\mu(t) = e^{-i H t} \mu e^{i H  t}$ is the system transition dipole moment in the Heisemberg picture. The system transition dipole operator $\mu$ is given by the sum of the dipole operators for the individual chromophores $\mu = \sum_{m} ( \mu_{m} |0 \rangle\langle m| + h.c. )$.
The dipole-dipole correlation function after transformation to the polaron frame reads
\begin{equation}
C_{d-d}(t) = \textrm{Tr}_{S+B} \{ \tilde{\mu}(t) \tilde{\mu}(0) \tilde{\rho}_{B} |0\rangle\langle 0| \}
\end{equation}
\noindent where the polaron frame dipole operator is given by:
\begin{eqnarray}
\tilde{\mu} &=& \sum_{m} \prod_{\mathbf{k}} \Big( D(-h_{\mathbf{k},m}) \mu_{m} |0 \rangle\langle m| + h.c. \Big) \nonumber\\
&=& \sum_{\alpha, m} \prod_{\mathbf{k}} \Big( D(-h_{\mathbf{k},m}) \mu_{m} u_{m\alpha} |0 \rangle\langle \alpha| + h.c. \Big)
\end{eqnarray}

\noindent where $D(h_{\mathbf{k},m})=e^{h_{\mathbf{k},m} b_{\mathbf{k}}^{\dagger} - h_{\mathbf{k},m }^{*} b_{\mathbf{k}}}$ is the bath displacement operator of mode $\mathbf{k}$ due to interaction with site $m$ and $u_{m\alpha}=\langle m | \alpha\rangle$ is the amplitude of site $m$ on renormalized eigenstate $|\alpha\rangle$.

Under the Markovian and secular approximations within the polaron frame, the dipole-dipole correlation function can be written as:
\begin{eqnarray}
C_{d-d}(t) =  \sum_{\alpha} C_{\alpha}(t) G_{\alpha}(t) \exp(i \omega_{\alpha} t)
\end{eqnarray}

\noindent For each renormalized exciton $|\alpha\rangle$ there are two contributions to the dipole-dipole correlation function: one arising from the displacement of the vibrational  environment, and a component due to dephasing of the electronic degrees of freedom within the polaron frame. The bath correlation function in the polaron frame $C_{\alpha}(t)$ is defined as

\begin{equation}
C_{\alpha}(t) = \sum_{mn} \mu_{m}^{*} \mu_{n} u_{m\alpha} u_{n\alpha} \textrm{Tr}_{B} \{ D(-h_{\mathbf{k},n}) D(h_{\mathbf{k},m}(t)) \tilde{\rho}_{B} \}
\end{equation}

\noindent The electronic correlation function is $G_{\alpha}(t) = \exp(-\tilde{\Gamma}_{\alpha}t)$ with a dephasing rate for each transition given by
\begin{equation}
\tilde{\Gamma}_{\alpha} = \sum_{\mu} \Big(\Gamma_{\alpha\mu,\mu\alpha}^{(1)} + \Gamma_{\mu\alpha\mu\alpha}^{(2)} + \Gamma_{\alpha\mu,\alpha\mu}^{(3)} + \Gamma_{\mu\alpha,\alpha\mu}^{(4)} \Big)
\end{equation}

\noindent and $\Gamma_{\alpha\beta,\mu\nu}^{(i)}$ Markovian polaron rates given in Kolli {et al.} \cite{kolli11}. Full details of the above derivation is presented in Appendix \ref{sec:a1}. 

The final form of the absorption spectra is given by
\begin{eqnarray}
A(\omega) = \sum_{\alpha} Re \int_{0}^{\infty} dt~e^{i(\omega-\omega_{\alpha}) t}~C_{\alpha}(t) G_{\alpha}(t)  \nonumber\\
\end{eqnarray}

In a similar manner we can derive the following expression for the fluorescence spectra:
\begin{eqnarray}
F(\omega) = \sum_{\alpha} n(\omega_{\alpha}) Re \int_{0}^{\infty} dt~e^{i(\omega-\omega_{\alpha}) t}~C^{*}_{\alpha}(t) G^{*}_{\alpha}(t)  \nonumber\\
\end{eqnarray}
\noindent where $n(\omega_{\alpha}) = (\exp(\beta \omega_{\alpha}) - 1)^{-1}$ is the thermal occupation of eigenstate $\alpha$.

\section{Signatures of quantized vibrations in spectra and dynamics of PE545}
\subsection{Signatures of quantized vibrations in the linear spectra}
\begin{figure}[t]
\begin{centering}
\subfloat[\scriptsize{PE545 absorption spectrum at room temperature (294 K)}]{
\includegraphics[width=86mm]{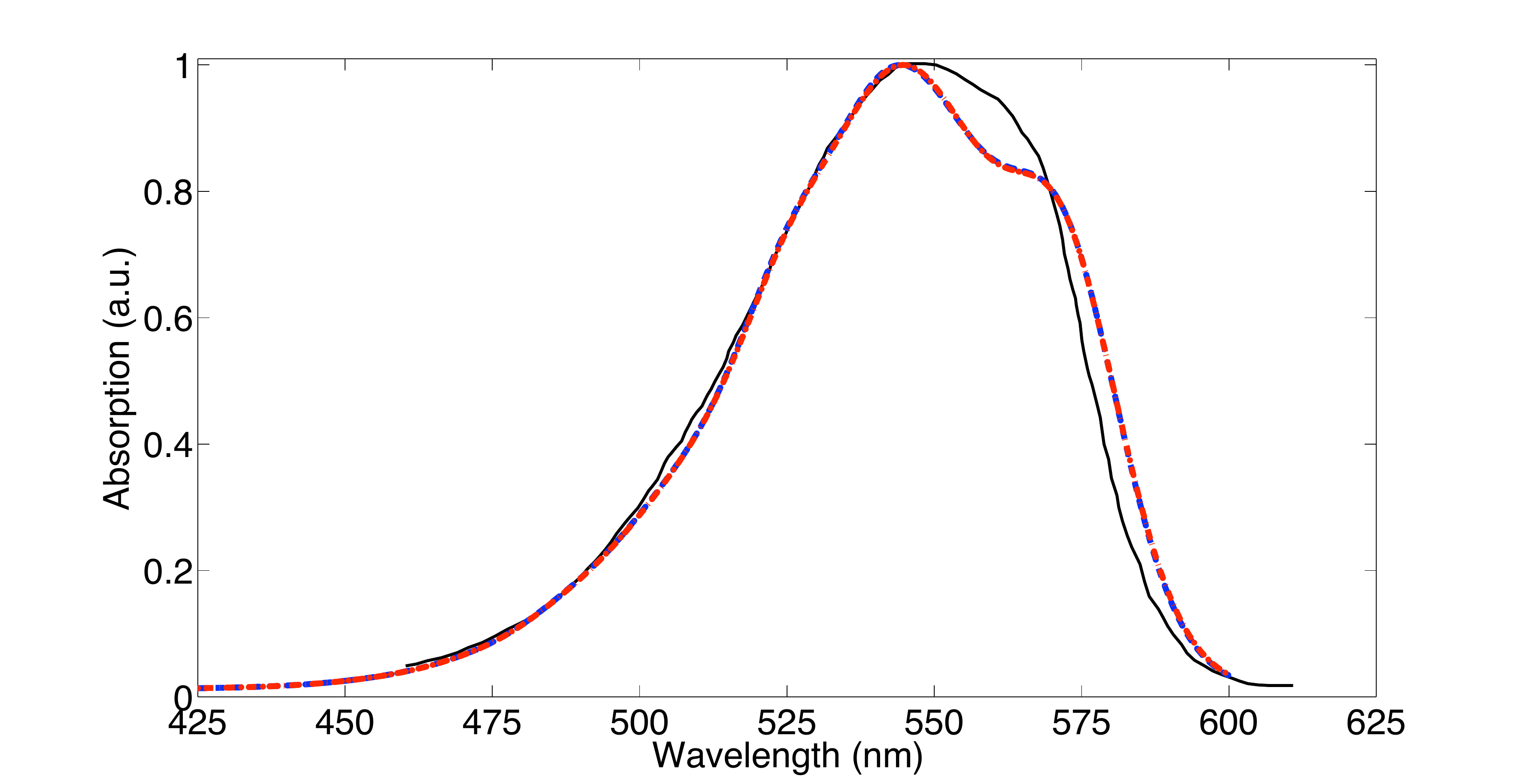}
} \\
\subfloat[\scriptsize{PE545 fluorescence spectrum at room temperature (294 K)}]{
\includegraphics[width=86mm]{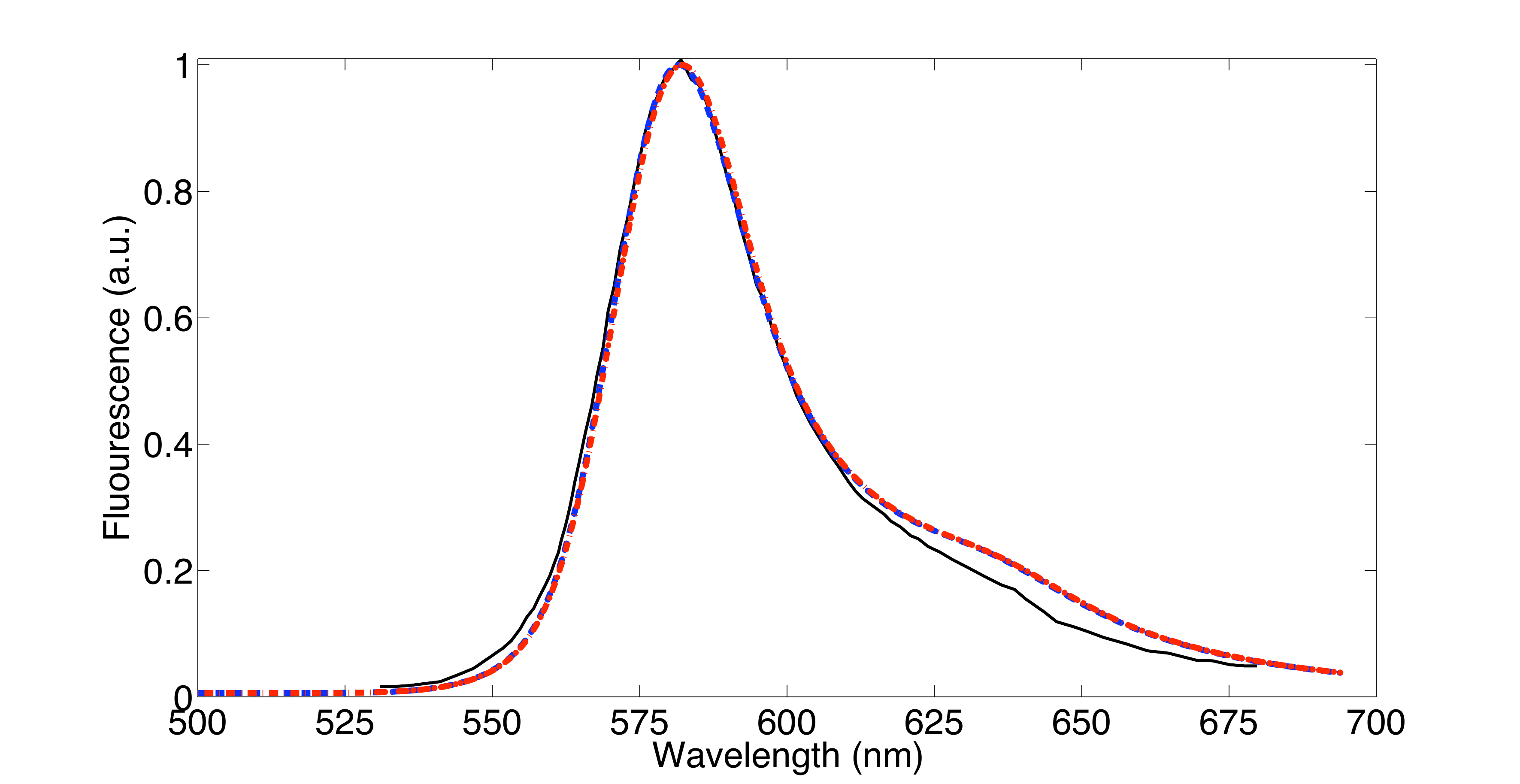}
}
\caption{Comparison of experimental (solid black) and calculated (solid red) spectra. Dashed-dotted curve indicates calculated spectra in the absence of solvent-induced correlations. The results presented include average over Gaussian static disorder for the site energies with FWHM$=500~\textrm{cm}^{-1}$ }
\label{fig:linearspectra}
\end{centering}
\end{figure}
Experimental signatures of specific vibrations participating in excitation dynamics in light-harvesting systems can be seen in photon-echo spectrocopy \cite{womick09}, resonance Raman studies \cite{west11}, fluorescence narrowing \cite{ratsep08, wendling00} and  hole-burning experiments \cite{raja93, jankoviak11}. Here we show that the theoretical framework considered reproduces known room-temperature spectra and we point out signatures of quantized high-energy vibrations in steady state fluorescence.

Absorption and fluorescence spectra are computed in the absence and presence of spatial correlations and averaged over 10000 realisations of Gaussian static disorder with FWHM$=500~\textrm{cm}^{-1}$. These results are presented in Figure \ref{fig:linearspectra} together with the experimental data of  Novoderezhkinin \textit{et al.} \cite{novoderezhkin10}. The polaron theory gives the correct location of the absorption peak frequency, though it does tend to underestimate the absorption just beyond this maximum (Figure \ref{fig:linearspectra}a). The differences in experimental and calculated absorption spectra may indicate that full quantum treatment of selected vibrations may be needed. Calculated and experimental fluorescence spectra are presented in  Figure \ref{fig:linearspectra}b. We recover the measured experimental peak fluorescence after shifting the theoretical result  by 6nm to the blue. Upon this correction, there is good qualitative agreement between theory and experiment.

Importantly, the main features of fluorescence are only recovered when quantized vibrations are included. While  the continuos component of the spectral density $J_{CM}(\omega)$ determines the width of the fluorescence profile, quantized high-energy vibrations described through $J_{QM}(\omega)$ give rise to the asymmetry observed towards the blue (around 630 nm).  The linear spectra are found to be unchanged as the degree of broadening, $b$, of the modes is varied from $1$ to $50~\textrm{cm}^{-1}$, being in good agreement with the results of Novoderezhkin \textit{et al.} \cite{novoderezhkin10}.

Finally, it is worth noting that we obtain near identical spectra in the absence and presence of spatial correlations. The PE545 complex is a system whose excitonic states remain highly localized for a wide range of spatial correlations. Consequently, the dipole-dipole correlation function and the resulting spectra are insensitive to the degree of the spatial correlations. In contrast to the linear spectra, excited state dynamics is susceptible to spatially correlated fluctuations as we discuss in the next section.
\subsection{Signatures of quantized vibrations in dynamics}
\begin{figure*}[]
\begin{center}
\subfloat[{Spatial distribution of excitation energy in presence of quantized vibrations.}]{
\includegraphics[width=120mm]{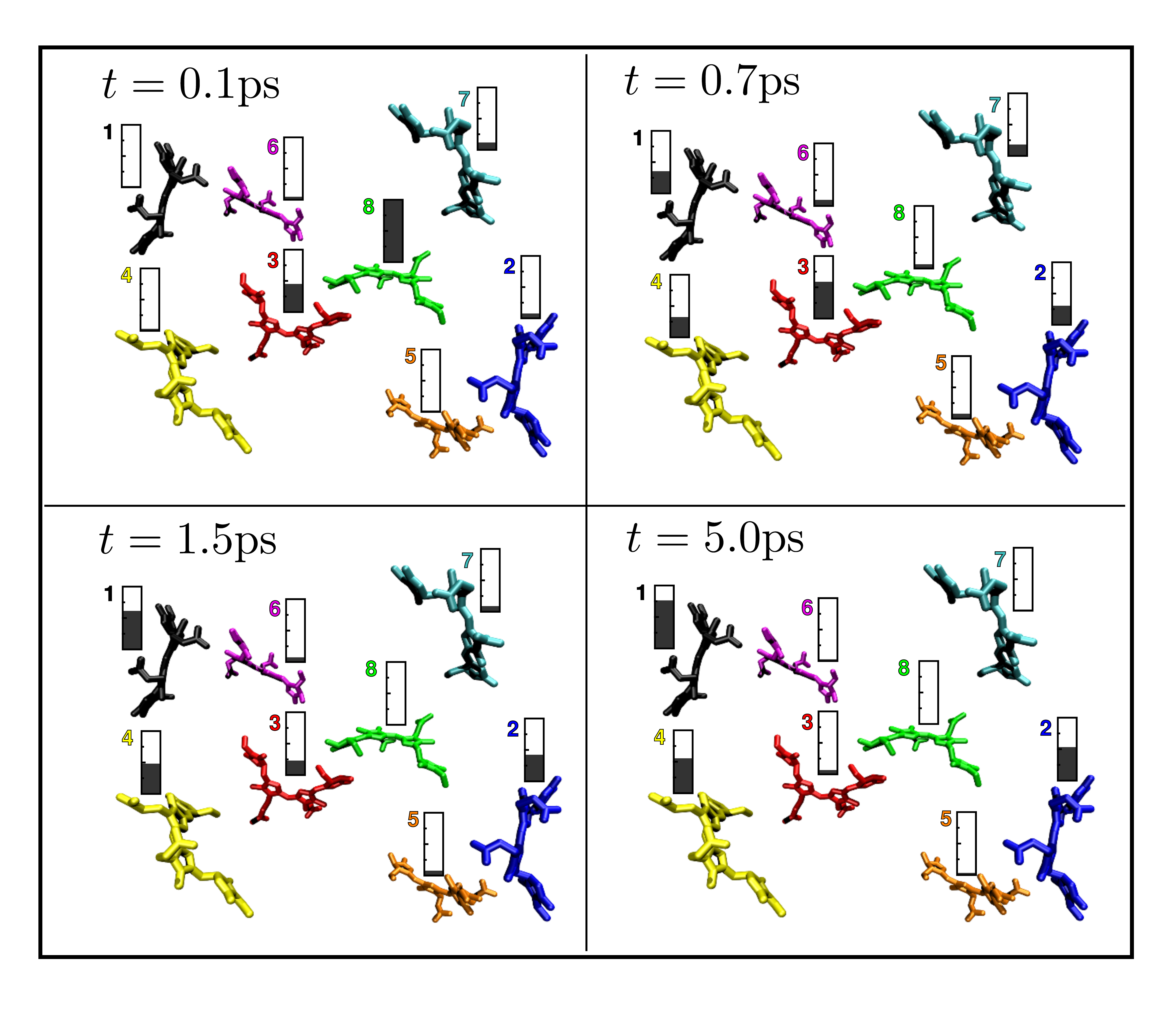}\label{fig:snapmodes}
}\\
\subfloat[{Spatial distribution of excitation in absence of quantized vibrations.}]{
\includegraphics[width=120mm]{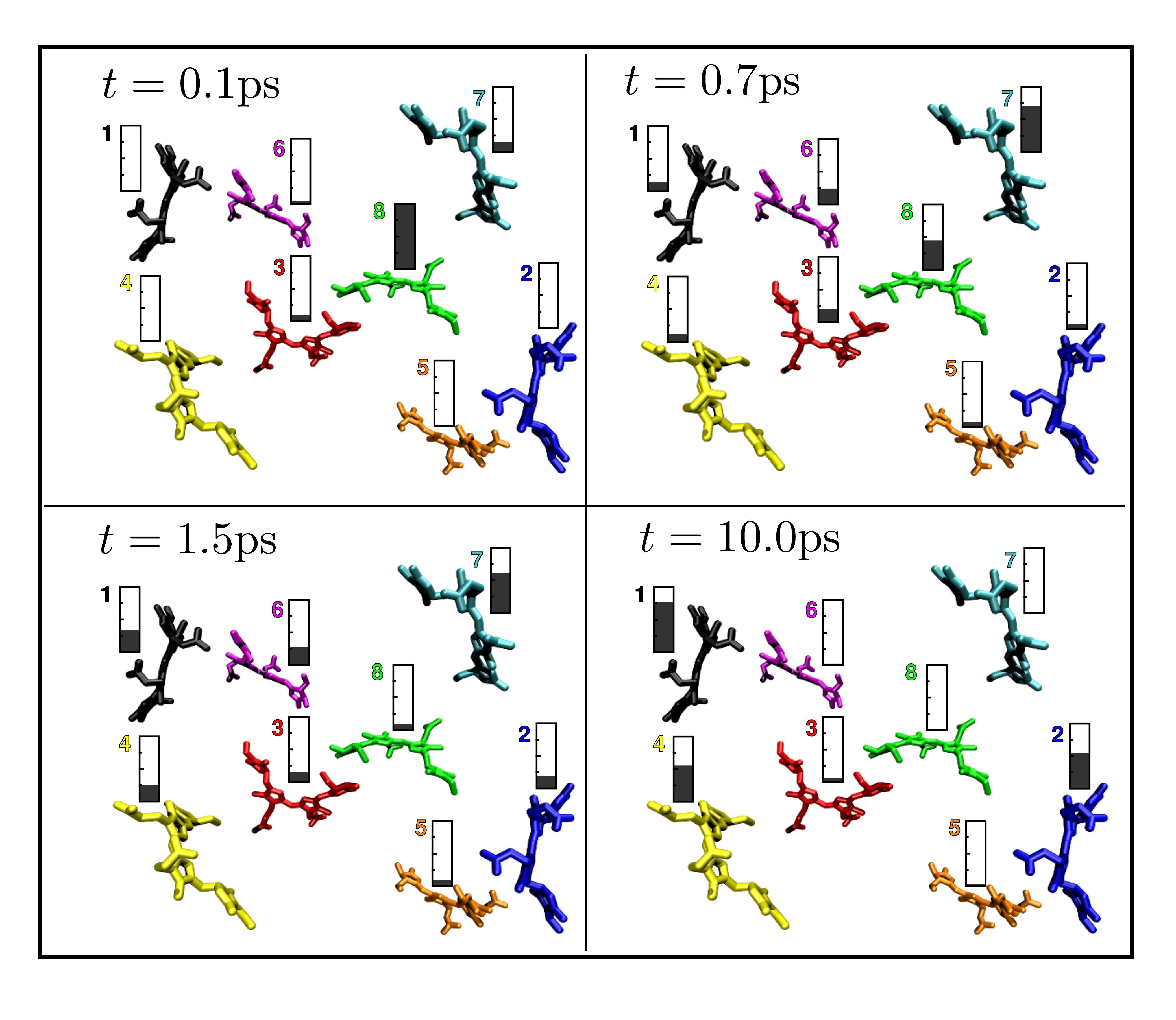}\label{fig:snapnomodes}
}
\caption{Snapshots of population dynamics. The filling of each bar is proportional to the probability of each site to be excited, with a maximum height denoting a population of 0.5 or above.}
\label{fig:snapshots}
\end{center}
\end{figure*}

\begin{figure*}[]
\begin{center}
\subfloat[Dynamics with quantized vibrations]{\includegraphics[width=85mm]{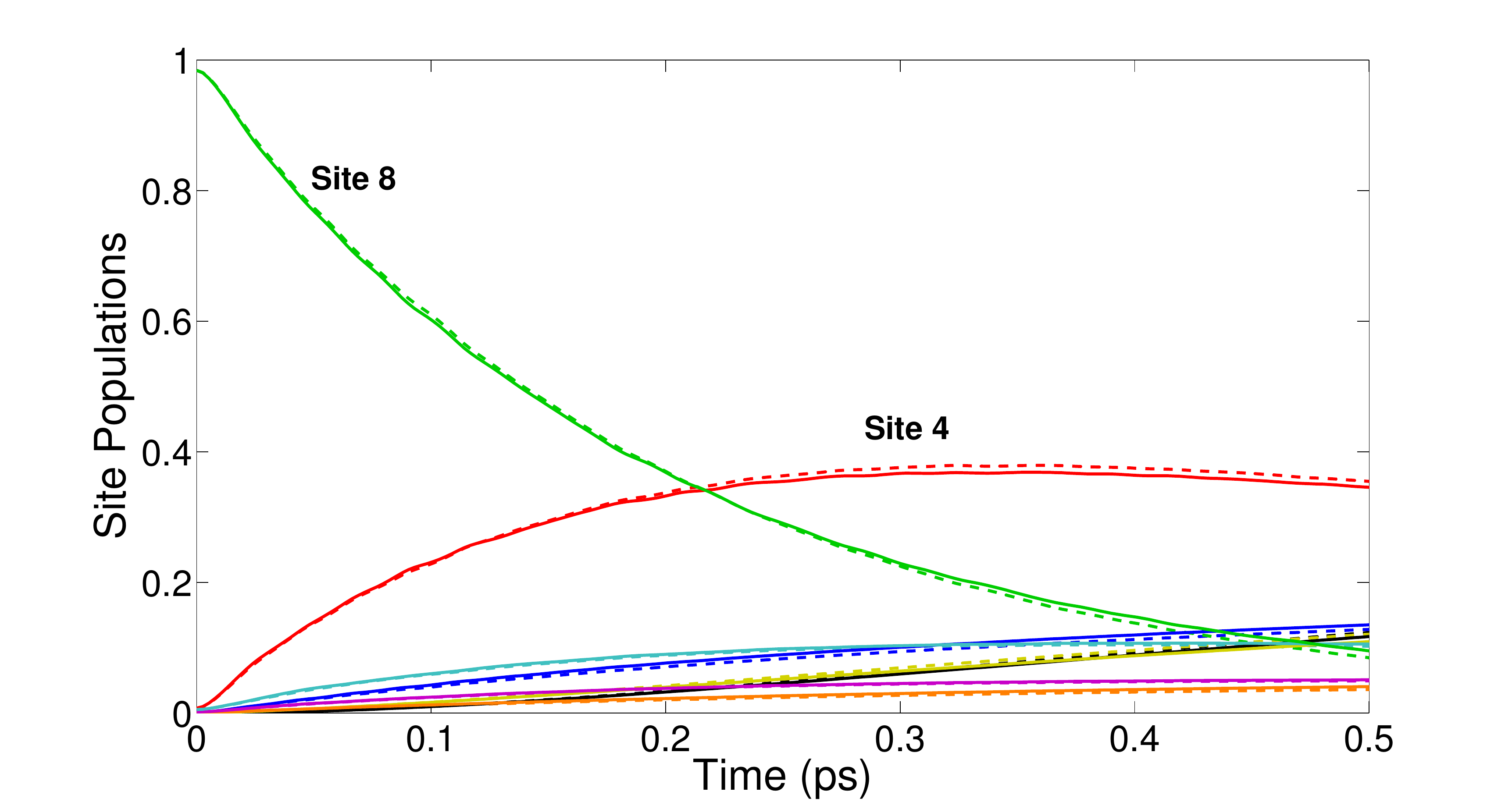}}
\label{fig:sitesmode}
\subfloat[Dynamics without quantized vibrations]{\includegraphics[width=85mm]{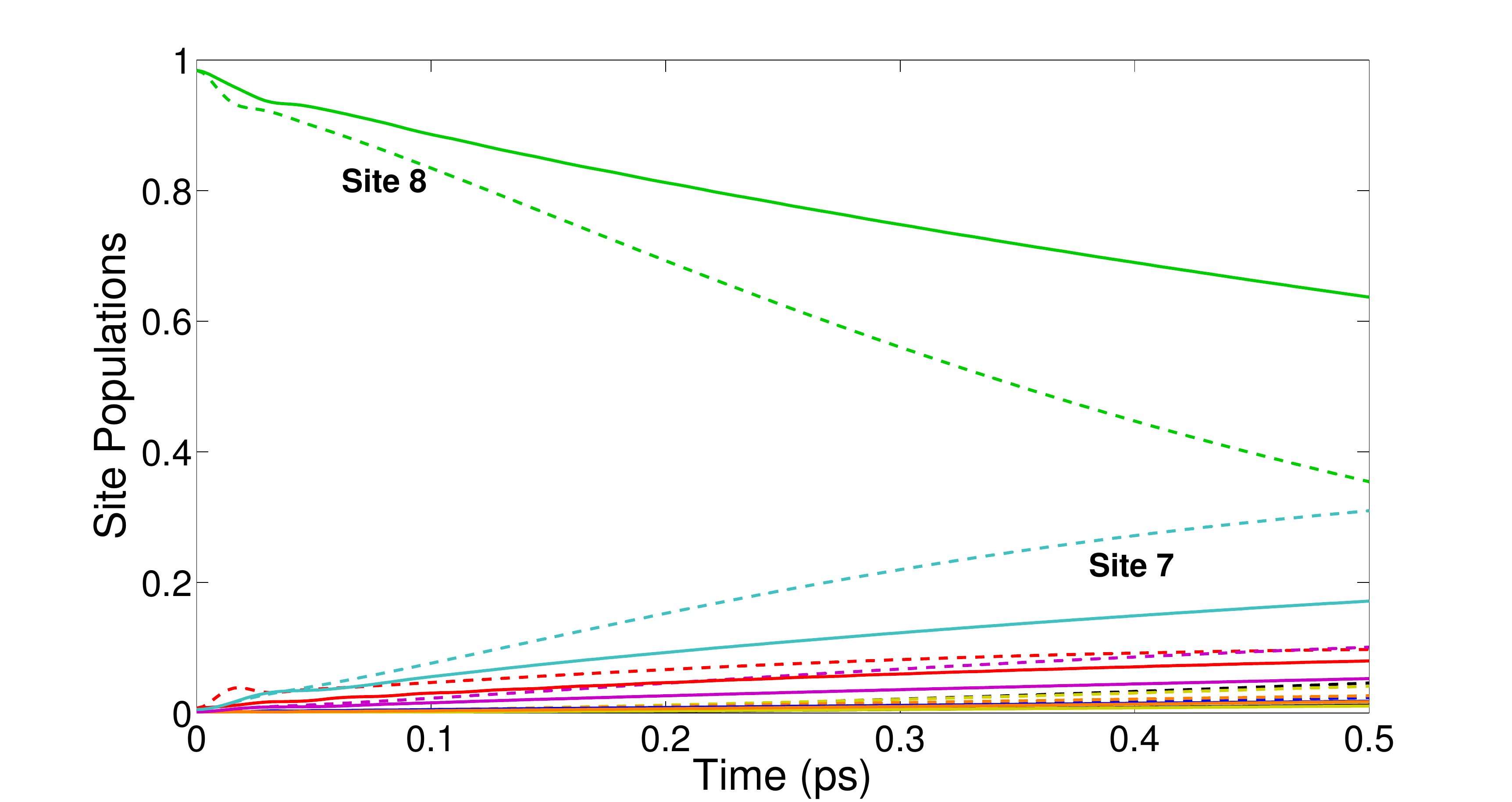}}
\label{fig:sitesnomode}
\caption{Comparison of site population dynamics with and without quantized vibrations. To make clear the differences between the two situations only two lines have been labeled. Dashed lines represent dynamics when solvent-induced correlated fluctuations are considered.}\label{fig:sitepopulations}
\end{center}
\end{figure*}
In this section we highlight the effects of quantized vibrations  by systematically comparing dynamics with and without such vibrations as well as comparisons when correlated fluctuations are included. We shall illustrate how  the presence of quantized vibrations is evidenced in transfer times and pathways of energy transport in PE545 and in the beating profile of populations and coherences. Before proceeding to the dynamics, however, let us first briefly discuss two important aspects regarding the initial state and the basis chosen to describe excitation dynamics. 

\textit{Initial state.-} We consider excitation by an ultrafast laser pulse in a time scale much smaller than the timescale of the solvation dynamics. Under these conditions, the excitonic system will undergo a vertical Franck-Condon transition where the vibrational degrees of freedom remains in thermal equilibrium during the fast excitation process. Therefore we may assume that the excitation pulse can excite an excitonic state $|e_m\rangle$ of the bare Hamiltonian (See Eq.(\ref{eq:hel})). In this work we assume that the laser pulse is shifted well into the blue end of the spectrum, such that it only excites the highest excitonic eigenstate of the  $|e_8\rangle$ corresponding to a highly localized excitation on $\rm{PEB_{50D}}$. This initial state is chosen in order to investigate the transfer pathways within the PE545 complex and to allow comparisons with previous works \cite{novoderezhkin10, collini10}. Furthermore this situation will let us evaluate whether coherent superpositions of excitonic states are evolved during the dynamics even if they were not present in the initial state. 

\textit{Basis of Relaxation.-}  It is worth taking a moment to consider the correct basis in which to describe the relaxation dynamics. Assuming the single-excitation system relaxes towards a quasi-equilibrium state, a natural basis of relaxation will be that in which the density matrix becomes diagonal as the steady state is approached. For arbitrary system-bath coupling, such a basis may be constructed in terms of the eigenstates of an effective Hamiltonian describing equilibration\cite{campisi09}. In practice, however, the evaluation of this effective Hamiltonian is not a straight forward task and therefore we must choose an approximate basis. Our calculations indicate that the long-time density matrix in the polaron frame is diagonal in the basis of the renormalized Hamiltonian Eq.(\ref{eq:heldressed}).  We therefore choose such a basis to describe relaxation but now in the untransformed frame.  In general, these renormalized states depend on the characteristics of system-bath interaction such that it will be different in the absence or presence of correlations. In the absence of solvent-induced correlations, we find that fluctuations induced by $J_{CM}(\omega)$ lead to  all bath-induced renormalisations $\beta_{mn}$ evaluating to zero. As a result, the master equation Eq.\ref{eq:TCL} becomes a set of non-Markovian rate equations describing incoherent excitation transfer between sites.  In the presence of spatial correlations corresponding to a speed of sound of $1000~\textrm{ms}^{-1}$, renormalisations are no longer zero and quantum coherent contributions to dynamics can be observed. The basis of relaxation in this case is a renormalized excitonic basis. 
\begin{figure*}[]
\begin{center}
\includegraphics[width=130mm]{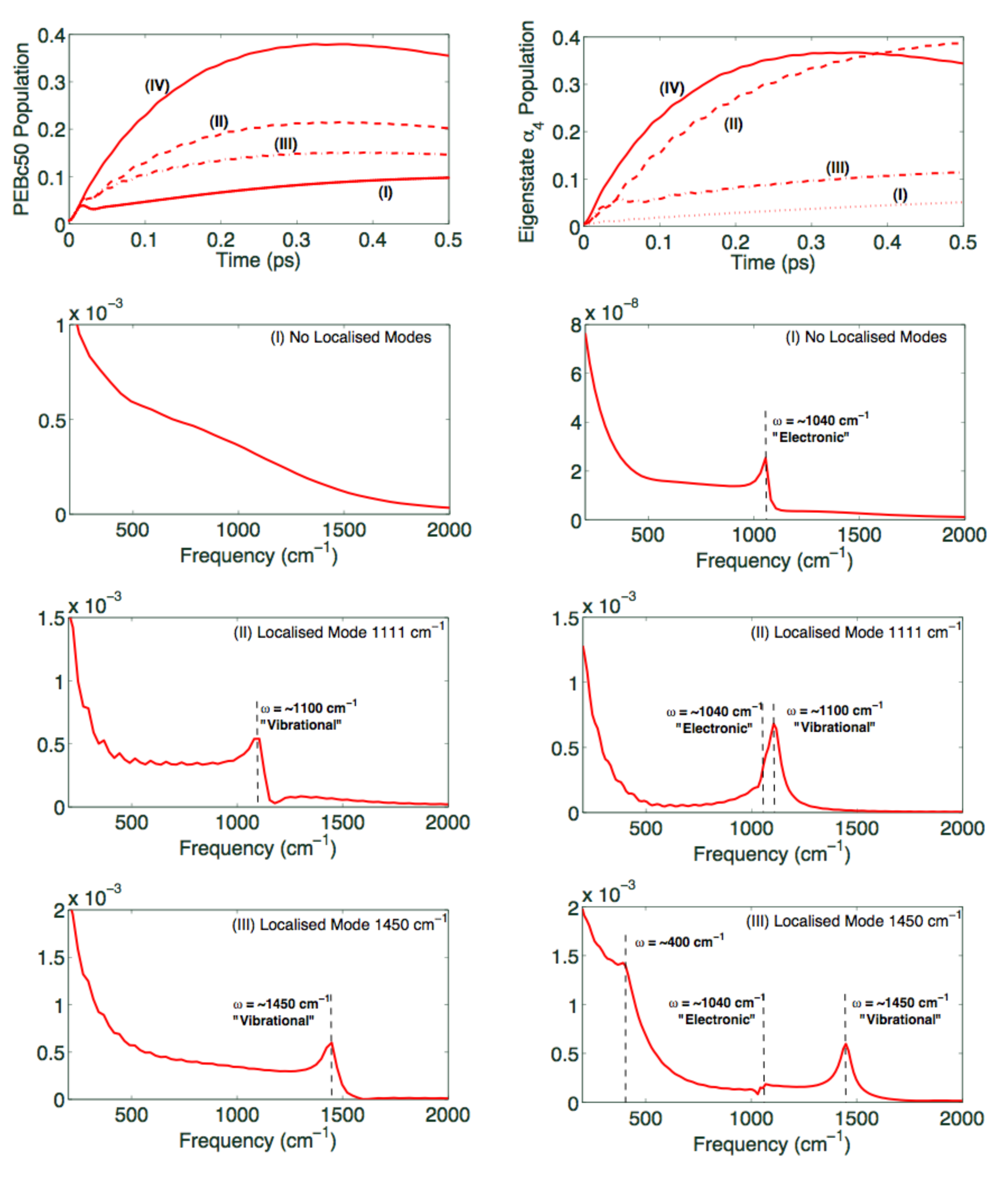}
\caption{Population dynamics of the excitonic state localized on $\rm{PEB_{50C}}$ in the absence of spatial correlations (left) and in the presence of solvent-induced correlations (right). Lines labeled with (IV) denote population when all modes are included. Panels (I) to (III) depict Fourier transforms of populations when no LM is included, or when only  $\omega_{QM}=1111\rm{cm}^{-1}$ or  $\omega_{QM}=1450\rm{cm}^{-1}$ are present. }
\label{fig:distributor}
\end{center}
\end{figure*}
\subsubsection{Transfer times and pathways}
Quantized vibrations drastically change the transfer times and pathways for excitation transfer in PE545. To illustrate this we begin by presenting in Figure \ref{fig:snapshots} snapshots of the distribution of the excitation in real space at different times in the presence of these vibrations (Figure \ref{fig:snapmodes}) and in their absence (Figure \ref{fig:snapnomodes}). The main feature that can be appreciated from these snapshots is that in both situations excitation energy initially localized on $\rm{PEB_{50D}}$ is finally distributed to the three lowest lying chromophores $\rm{PEB_{82C}}$, $\rm{DBV_{A}}$ and $\rm{DBV_B}$. However, the timescale of this process is 5 ps when assisted by quantized vibrations and becomes 10 ps if they are neglected. Importantly, the timescale observed in experiments \cite{novoderezhkin10}  is about 5 ps agreeing with our predictions. 

Comparison of the snapshots at 0.1 ps indicates that quantized vibrations \textit{activate transfer} to its nearest neighbour chromophore 
$\rm{PEB_{50C}}$ (also the most strongly interacting site). This can also be appreciated in Figure \ref{fig:sitepopulations}a depicting site population dynamics as a function of time. In contrast,  when no quantized vibrations are included, excitation is distributed to the chromophore nearest in energy 
$\rm{PEB_{82D}}$ as can be seen in Figure \ref{fig:sitepopulations}b).  One striking feature of the distribution profile at 0.7~ps in Figure \ref{fig:snapmodes} is that excitation energy has been widely distributed over all the complex, while in the absence of fast vibrations energy is concentrated mainly on protein subunit D. Again, results at these intermediate timescales are in good agreement with experimental observations in \cite{novoderezhkin10}. The features observed in the presence of fast vibrations allow PE545 to perform its biological function: distribution of excitation across the antenna implies that energy can be efficiently transferred to other nearby antennae, and fast transfer to units A and B, which are those probably in contact with PSI or PSII \cite{wit06}, will aid energy conversion. This is further discussed in section \ref{sec:role}. Finally, it is worth nothing that the participation of quantized vibrations in excitation dynamics guarantees that the general timescales of transfer and overall distribution of excitation energy remains practically unchanged as correlated fluctuations are considered (Figure \ref{fig:sitepopulations}a). This is no longer the case when fast modes are excluded. In this situation, since transport is mostly driven by thermal fluctuations, a small degree of correlations slows transfer (Figure \ref{fig:sitepopulations}b). The interplay between correlated fluctuations and quantized vibrations does exhibit interesting features when contributions of each individual mode to site populations are analyzed as shown in the next section.

\subsubsection{Population dynamics}
High-energy vibrations quasi-resonant with excitonic transitions manifest themselves not only in the activation of transport between widely energetically separate states but also in the oscillatory pattern of the dynamics. This arises primarily from the non-equilibrium dynamics of the vibrational components. To illustrate these features, we focus now on the population evolution of of the renormalized state $|\alpha_4\rangle$, which is mostly localized on the central pigment $\rm{PEB_{50C}}$.  Figure \ref{fig:distributor}(left) presents the population of $\rm{PEB_{50C}}$ when the 14 vibrational modes are included and also when only the modes with frequencies $\omega_{QM}=1111~\textrm{cm}^{-1}$ or $\omega_{QM}=1450~\textrm{cm}^{-1}$ are present. While $\omega_{QM}=1111~\textrm{cm}^{-1}$ is in close resonance with the energy detuning between the initially excited site and the central site, the mode with $\omega_{QM}=1450~\textrm{cm}^{-1}$ has the largest coupling to each site. We find that both modes are able to activate  transport in comparison to the situation when no mode is included (curve labeled (I)), with a larger rate enhancement achieved with the quasi-resonant mode as expected. The Fourier spectrum of the population dynamics with the full set of 14 modes is somewhat complicated (not shown). Hence we investigate the oscillatory behaviour when only one mode is present (see (I) to (III) in Figure \ref{fig:distributor}(left)).  These figures show that the subtle oscillations of these populations are due to the corresponding vibrational modes. 

It is important to point out that off-resonance modes such as $\omega_{QM}=1450~\textrm{cm}^{-1}$ are able to activate transport  to $\rm{PEB_{50C}}$ due to the interplay with thermal-induced fluctuations induced by $J_{CM}(\omega)$: fluctuations in excitonic energy splittings cause excitonic transitions to be close to an otherwise off-resonance vibration.  Equally, fluctuations bring out of resonance transitions which are otherwise quasi-resonant with a fast mode. The interplay between thermal fluctuations and quantized vibrations bring us to the discussion of the interesting behaviour observed when a small degree of correlated  fluctuations is included (see Figure \ref{fig:distributor}(right)). In this case, excitonic transitions remain quasi-resonant with a particular mode that fully enhances transport as shown in Figure \ref{fig:distributor}(right) (curve labeled (II)). Furthermore, the population dynamics exhibits frequency components of both excitonic and vibrational origin. This can be seen in Figures \ref{fig:distributor}(right)(I)-(III)  which show the Fourier spectrum of populations when there is a small degree of solvent-induced correlations. When only the mode with $\omega_{QM}=1450~\textrm{cm}^{-1}$ is included the spectrum exhibits three peaks. One peak at $\sim1450~\textrm{cm}^{-1}$ due to the localized vibration assisting transport and a low-amplitude peak at $\sim1040~\textrm{cm}^{-1}$ which compares very well with the energy difference between eigensates $|\alpha_8\rangle$ and $|\alpha_4\rangle$. Hence we may attribute this frequency component to a quantum-coherent excitonic oscilation and indicates that, in the basis chosen to describe the dynamics, populations can witness coherences.  The feature at $400~\textrm{cm}^{-1}$ in Figures \ref{fig:distributor}(III)(right) does not correspond exactly to any transition frequency, therefore we believe that this peak arises from a beating between a frequency of excitonic coherence ( $\omega\sim 1040~\textrm{cm}^{-1}$) and the vibration-induced oscillation of frequency $\omega_{QM}=1450~\textrm{cm}^{-1}$. For comparison, we present the Fourier spectrum when no mode is present (see Figures \ref{fig:distributor}(right)(I)) and when only the mode $\omega_{QM}=1111~\textrm{cm}^{-1}$ is included (Figures \ref{fig:distributor}(II)(right)). The small scaling factor of the spectra in Figure \ref{fig:distributor}(right)(I) is consistent with the very small amplitude of the population oscillations observed when vibrational modes are not present. 

\subsubsection{Coherence dynamics}
\begin{figure}[]
\begin{center}
\subfloat[\scriptsize{Real parts of $\rho_{\alpha 8,\alpha 7}$ (thick blue) and $\rho_{\alpha 8,\alpha 6}$ (thin green)}]{
\includegraphics[width=86mm]{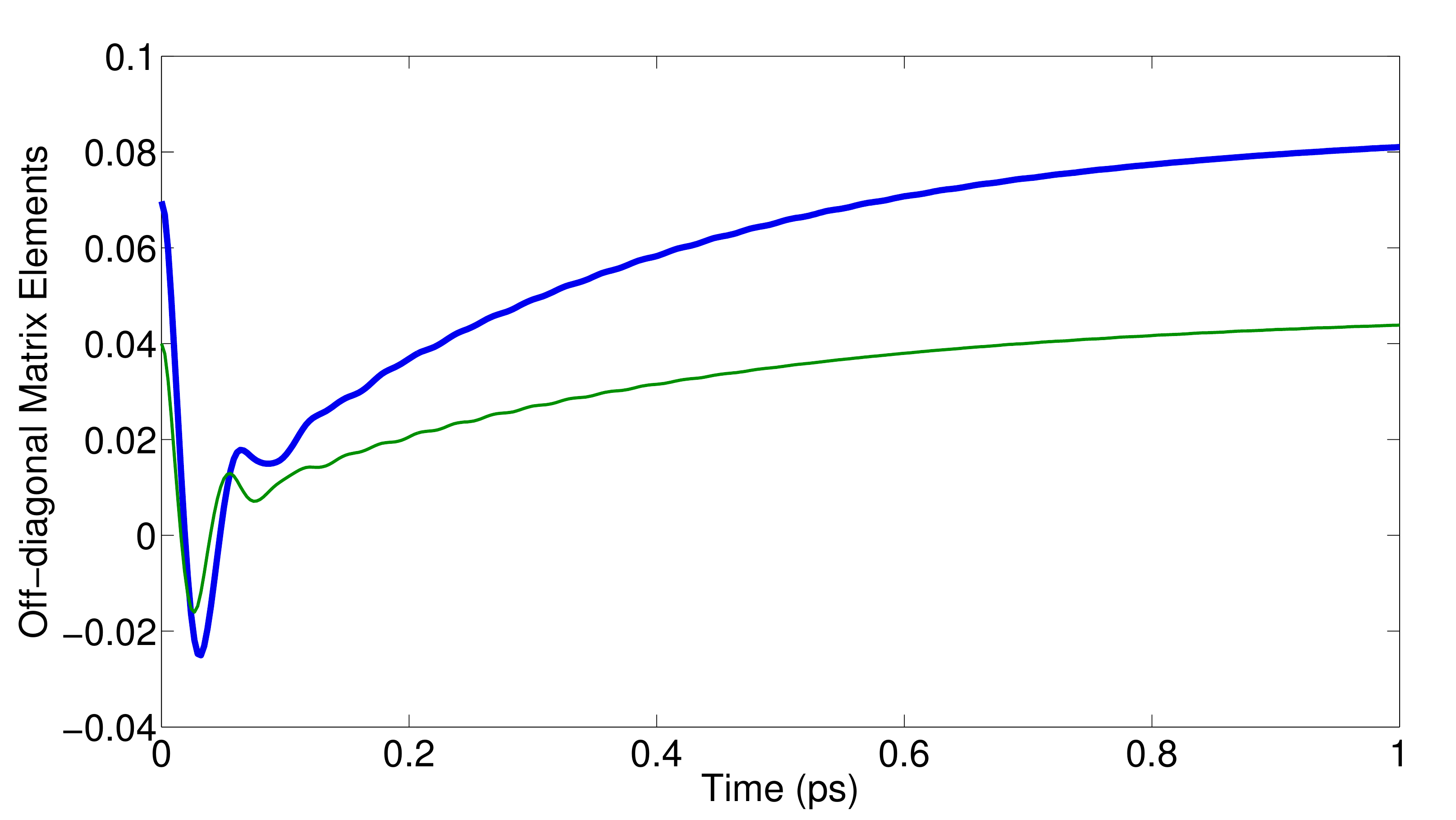}
\label{fig:coherences}
} \\
\subfloat[\scriptsize{FT of Re$\{\rho_{\alpha8,\alpha7}\}$ (thick blue) and Re$\{\rho_{\alpha 8,\alpha 6}\}$ (thin green) over the time interval 0 to $1~\rm{ps}$.}]{
\includegraphics[width=86mm]{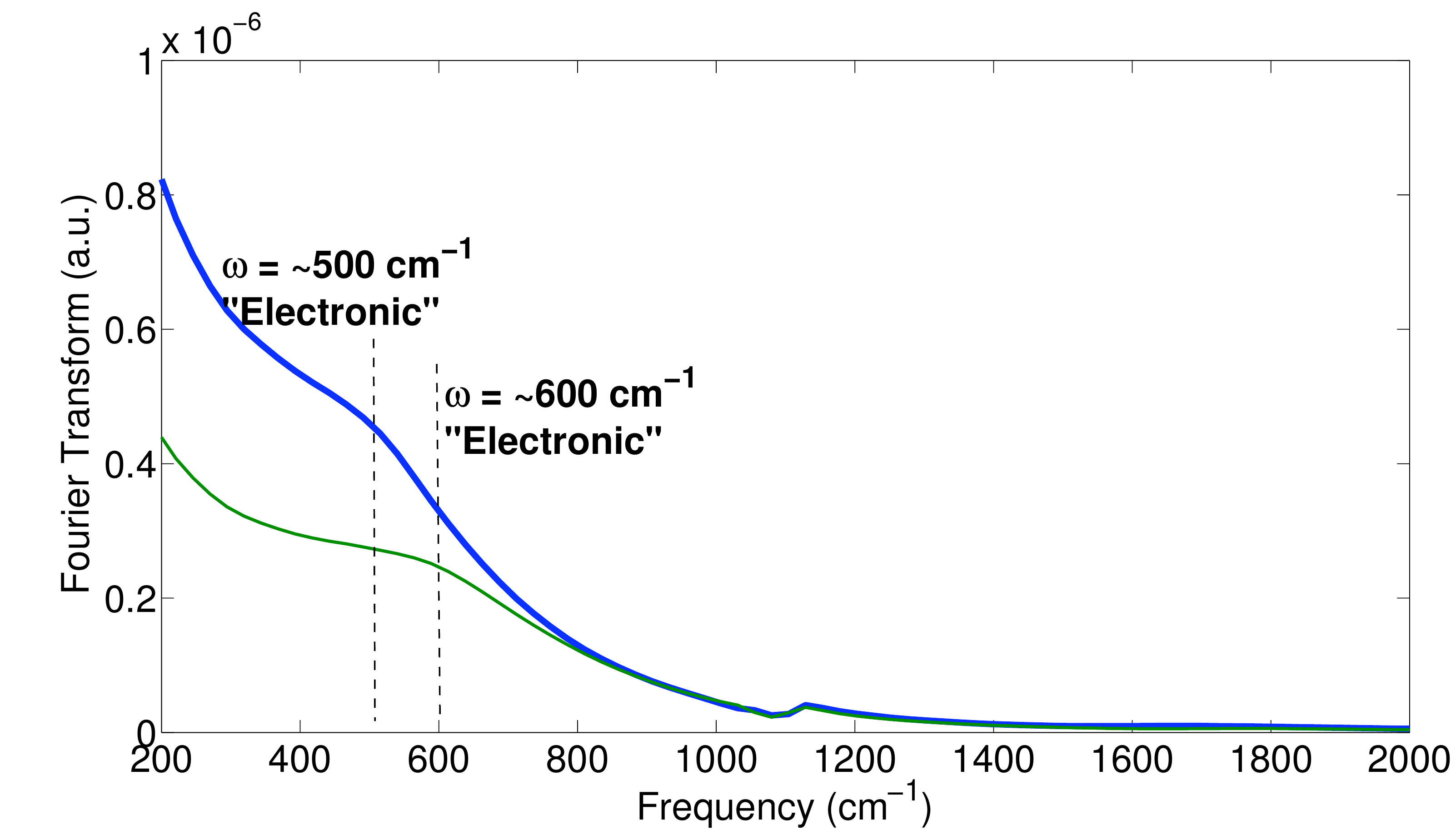}
\label{fig:coherencesFT}
} \\
\subfloat[\scriptsize{FT of Re$\{\rho_{\alpha 8,\alpha 7}\}$ (thick blue) and Re$\{\rho_{\alpha 8,\alpha 6}\}$ (thin green) over time interval of $250~\textrm{fs}$ to $1~\textrm{ps}$.}]{
\includegraphics[width=86mm]{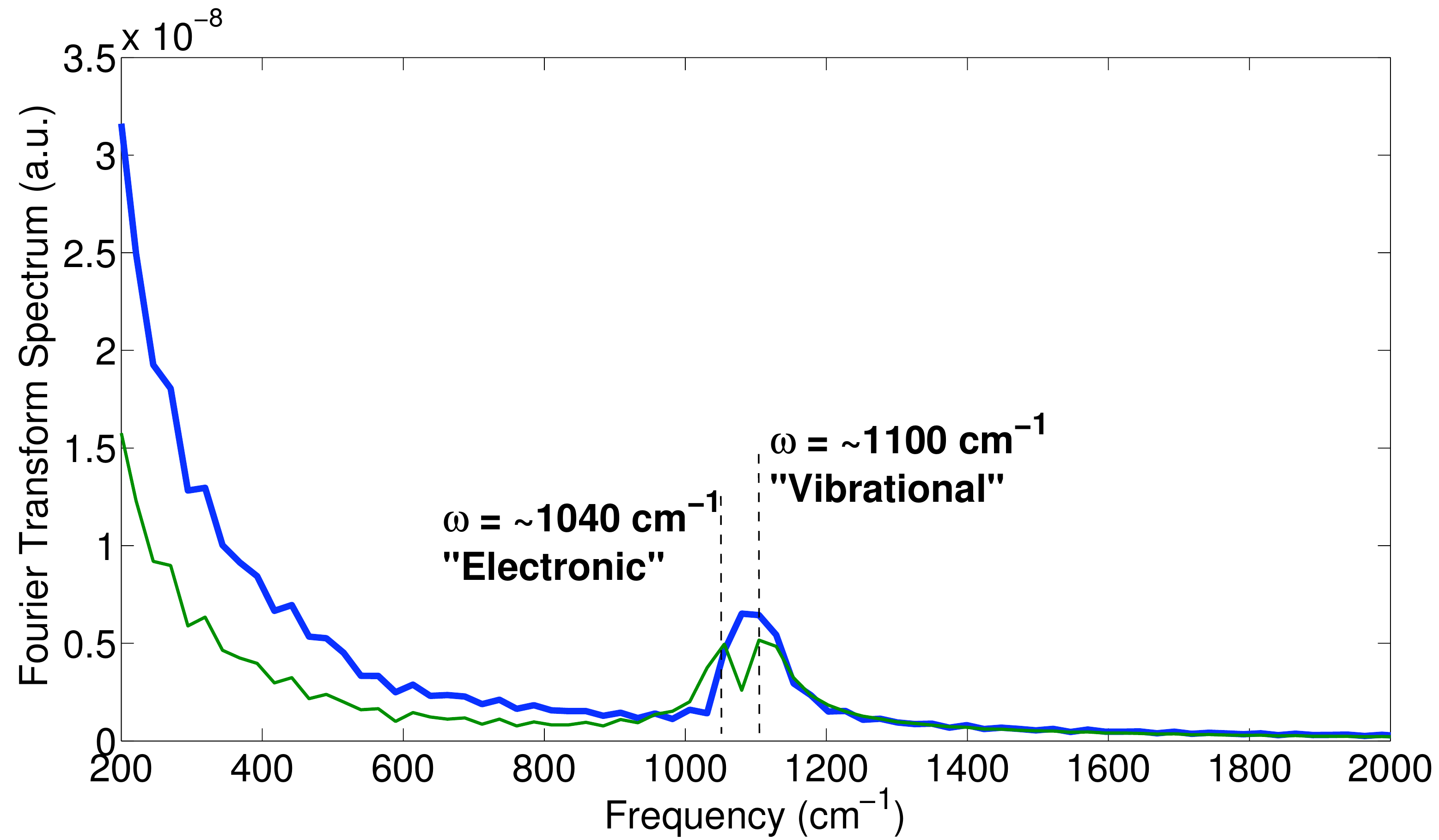}
\label{fig:coherencesFT2}
}
\caption{Excitonic coherences present when spatial correlations are included and their frequency spectrums}
\label{fig:distributercorrelations}
\end{center}
\end{figure}
As discussed in the previous section, quantum-coherent contributions to dynamics appear when solvent-induced correlated fluctuations are included. In this situation, bath-induced renormalizations do not vanish and they range between $\beta_{28}\sim 0.11$ and $\beta_{45}\sim 10^{-5}$ such that quantum coherent dynamics associated to $\tilde{H}_{el}$ is at play. We therefore now investigate how non-equilibrium vibrations manifest themselves in the oscillatory profile of excitonic coherences. For simplicity, we consider the situation where only the mode $\omega_{QM}=1111~\textrm{cm}^{-1}$ is present and show in Fig. \ref{fig:coherences}  the evolution of the coherences between the two highest energy renomalised eigenstates i.e. $\rho_{\alpha 8,\alpha 7}=\langle \alpha_8|\rho(t)|\alpha_7\rangle$ as well as $\rho_{\alpha 8,\alpha 6}=\langle \alpha_8|\rho(t)|\alpha_6\rangle$. These off-diagonal matrix elements correspond to coherences investigated in room-temperature two-dimensional spectroscopy \cite{collini10}. 

Large-amplitude oscillations are observed in the dynamics of both $\rho_{\alpha 8,\alpha 7}$ and $\rho_{\alpha 8,\alpha 6}$ but they last only for about 50-100 fs. On longer timescales, there is also a small-amplitude oscillatory component -a hundred times smaller than the oscillations in the short time scale. To investigate the origin of the two different oscillatory features, we once again calculate their Fourier transforms. The frequency spectrums over the 1~ps timescale are presented in Figure \ref{fig:coherencesFT}. Two features should be noticed. First, there is a  very broad component due to the fast decaying oscillations seen at short times. For the matrix element $\rho_{\alpha8,\alpha7}$ this broad feature occurs at $500~\textrm{cm}^{-1}$, which corresponds to the energy difference between eigenstates $|\alpha_8\rangle$ and $|\alpha_7\rangle$. Similarly for $\rho_{\alpha8,\alpha6}$ this broad features is observed around $\omega=600~\textrm{cm}^{-1}$, which corresponds to the energy difference between eigenstates $|\alpha8\rangle$ and $|\alpha7\rangle$. Therefore, we may attribute these frequencies to quantum coherent evolution associated to $\tilde{H}_{el}$. In contrast, the second feature within the Fourier spectra is much subtler and occurs in both cases at roughly $1100~\textrm{cm}^{-1}$. 

To investigate the fine structure of the spectrum of long-lived oscillations we perform the Fourier transform only over the time interval $250~\textrm{fs}$ to $1~\textrm{ps}$ (Figure \ref{fig:coherencesFT2} ). Both off-diagonal elements exhibit a peak at $1100~\textrm{cm}^{-1}$, which can be attributed  to vibration-induced oscillations due to the localized mode with $\omega_{QM}=1111~\textrm{cm}^{-1}$. Interestingly, however, the matrix element $\rho_{\alpha 8,\alpha 6}$ exhibits an additional peak at $1040~\textrm{cm}^{-1}$, an energy scale comparable to the transition frequency between eigenstates $| \alpha_8\rangle$ and $|\alpha_4\rangle$. This is an indication of coherence-coherence coupling during the evolution and shows that excitonic coherences are generated during the dynamics even if they were not initially created. 

In summary, we have shown that in the presence of a small degree of correlations supporting coherent evolution of electronic excitations, the beating pattern of excitonic coherences in the presence of vibrations out of equilibrium has a dual origin: short-time oscillations lasting for about 50-100 fs reflecting the time scale associated to the electronic Hamiltonian and a superimposed vibration-induced modulation that survives close to the picosecond timescale. Importantly, the two timescales identified here compare very well with recent experimental results on a similar complex \cite{turner12}.

\section{Role of quantized vibrations}
\label{sec:role}
\begin{figure}[]
\begin{center}
\subfloat[\scriptsize{Non-cascaded energy transfer between protein subunits when quantized vibrations are included}]{
\includegraphics[width=85mm]{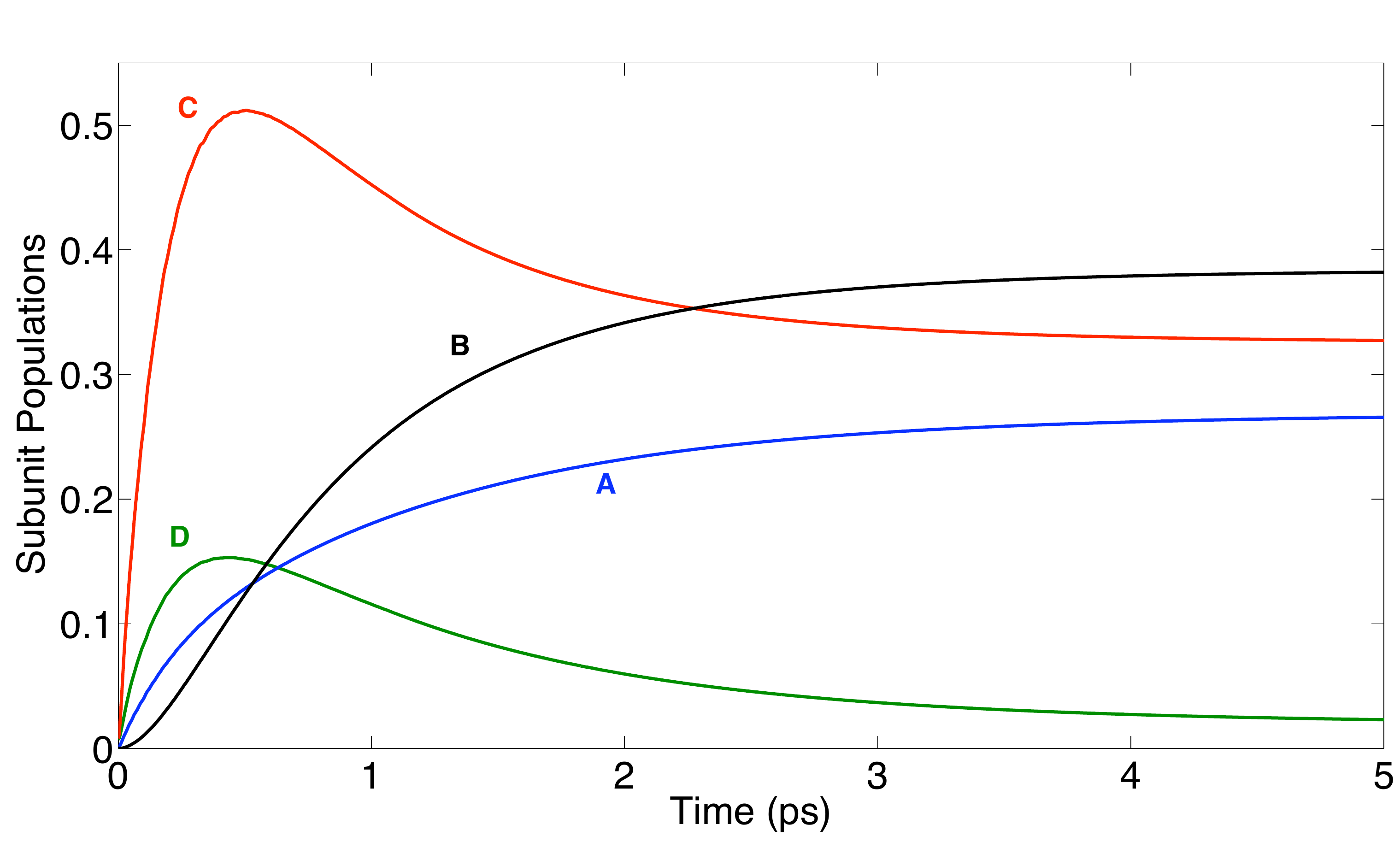} \label{fig:noncascaded}
}\\
\subfloat[\scriptsize{Cascaded energy transfer from protein subunit D to A when quantized vibrations are neglected}]{
\includegraphics[width=85mm]{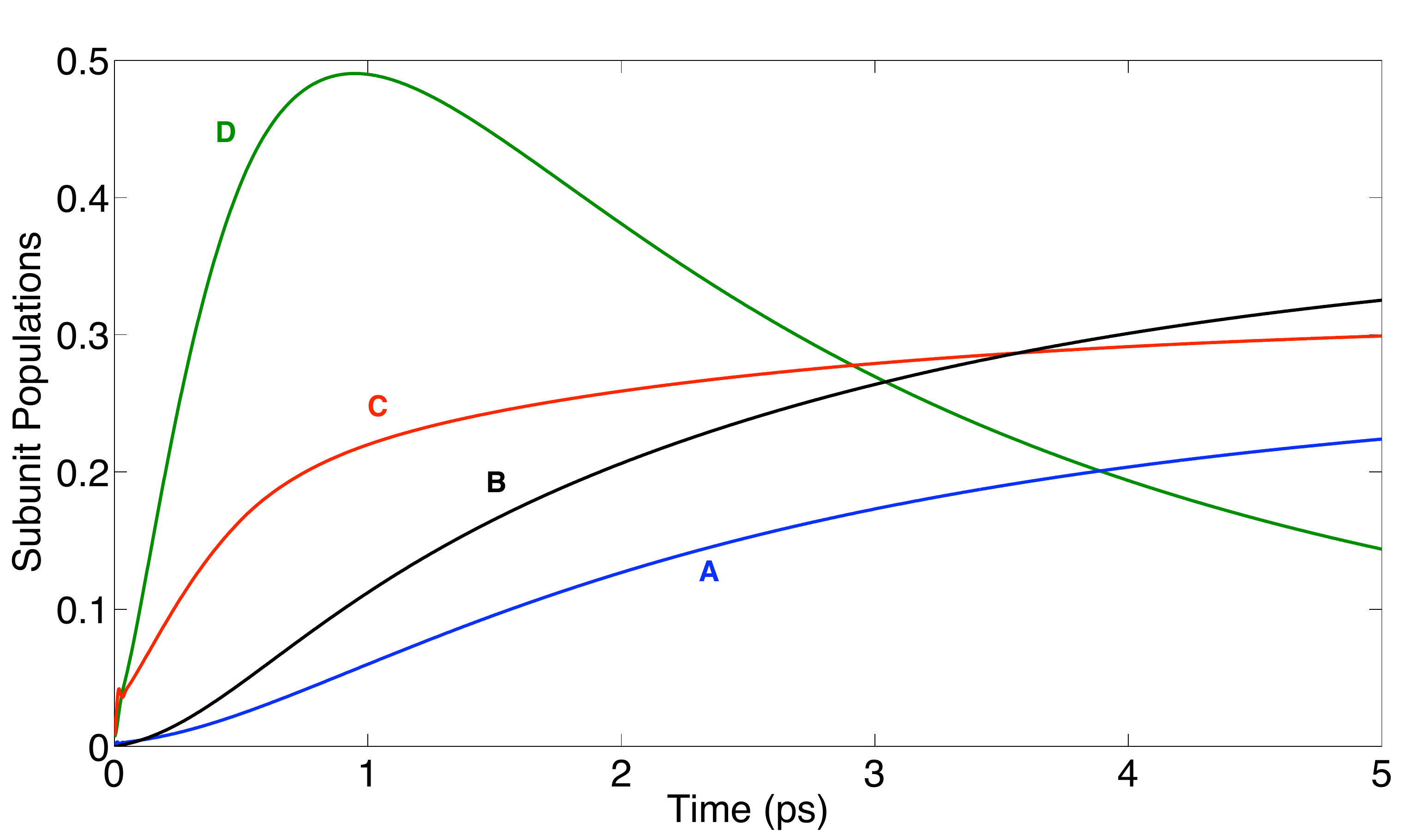}\label{fig:cascaded}
}
\caption{Total population within each protein subunit of PE545. For subunit D we have neglected the contribution of the initially excited chromophore $\rm{PEB_{50D}}$.}
\label{fig:subunitpopulations}
\end{center}
\end{figure}

\begin{figure}[]
\begin{center}
\subfloat[\scriptsize{Probability of excitation transfer to acceptor states}]{
\includegraphics[width=85mm]{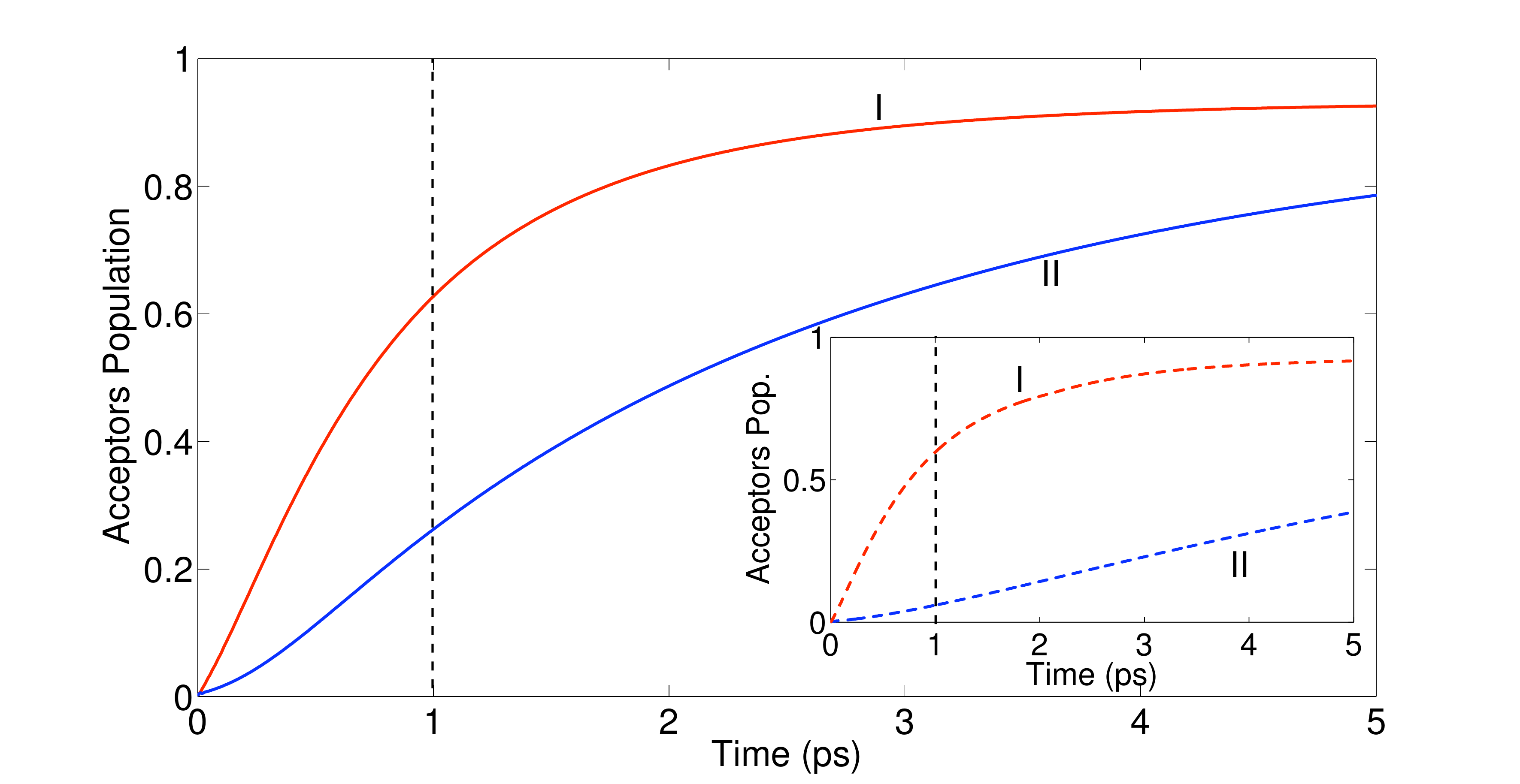}
\label{fig:temporaldistribution}
}\\
\subfloat[\scriptsize{Spatial spread of excitation energy}]{
\includegraphics[width=85mm]{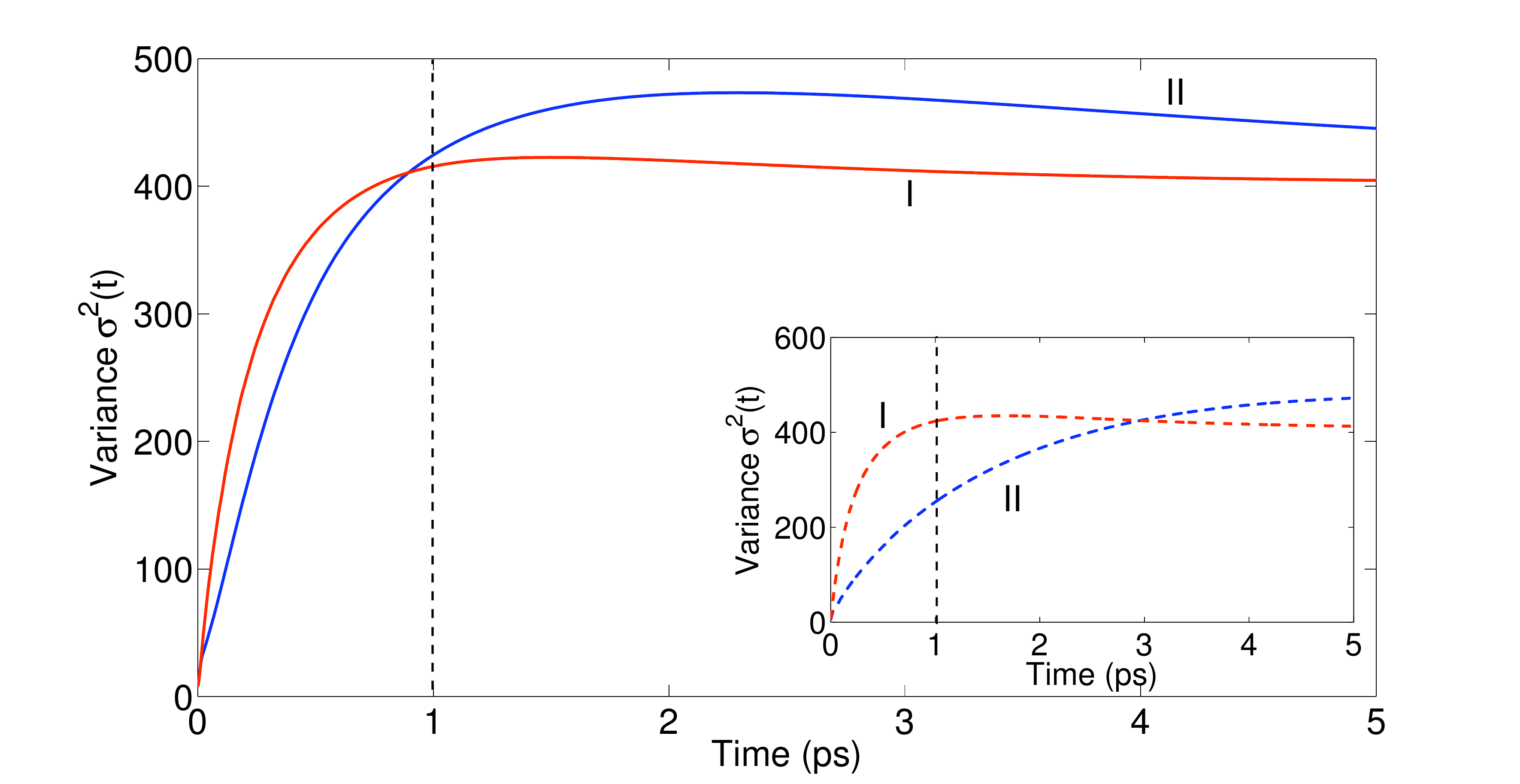}
\label{fig:spatialdistribution}
}
\caption{Performance measures in the presence (I) and absence (II) of quantized vibrations. Insets illustrate the behaviour when solvent correlations are considered. The dotted line at $1~\rm{ps}$ indicates the estimated transfer time between antennae}
\label{fig:temporalspatialdistribution}
\end{center}
\end{figure}
The results of the  preceding section allow us to clearly point out three key roles of the strong-coupling between highly localized electronic excitations and quantized vibrations in the biological function of PE545. 

\subsection{Promotion of non-cascaded energy transfer}
In the absence of quantized vibrations exciton-phonon interactions are dominated by the continuous distribution of modes that induce transitions between states near in energy. Given the local nature of excitonic states, transfer dynamics follows the hierarchy of energies of the chromophores. In light of the intimate connection between the hierarchy of energies and the physical arrangement of the pigments in the protein subunits (Figure \ref{fig:table1}), we conclude that when quantized vibrations are neglected, electronic excitation moves through PE545 in a sequential manner, from D onto subunit C and finally to subunits B and A as shown in Figure \ref{fig:cascaded}. Clearly this would be a very unfavourable dynamics for the biological function of transferring energy to other antennae.
 
In contrast, quantized vibrations take advantage of electronic interactions to promote non-sequential transfer by bridging the energy gap between subunits D and C and spreading out energy from each of them to A and B as shown in Figure \ref{fig:noncascaded}. In short, {\it quantized vibrations in close resonance with excitonic transitions provide a built-in mechanism to control energy transfer pathways}.  In general, explotation of environment and system resonances is an important design principle for directing energy transfer \cite{perdomo10}. 

\subsection {Optimal spatial and temporal spread of excitation} 
High-energy vibrations in PE545 guarantee an effective distribution of energy throughout the complex towards the acceptor states. The activation of energy transport between $\rm{PEB_{50D}}$ and $\rm{PEB_{50C}}$ generates both wider spatial spread of excitation to allow transfer between antennae and also faster transfer to the final acceptor states (localized on $\rm{DVB_{A}}$, $\rm{DVB_{B}}$ and $\rm{PEB_{82C}}$) to guarantee energy conversion if an PE545 antennae is in close proximity to complexes like the PSII or PSI. 

Quantitative support of the above statement is shown in Figure \ref{fig:temporalspatialdistribution} which summarizes the dual role of the quantized vibrations  in promoting \textit {faster transfer and wider spatial distribution of excitation energy}. The total population of all the acceptor states is depicted in Figure \ref{fig:temporaldistribution} and clearly shows a significant enhancement of the transfer rate in the presence of promoting vibrations.  To quantify the spatial spreading of excitation energy in PE545 complex we compute the variance $\sigma^{2} (t)$ in the spatial position of the excitation defined as follows:
\begin{equation}
\sigma^{2}(t) = \sum_{m} p_{m}(t) \big(|\mathbf{r}_{m}|^{2} - r^{2}(t) \big),
\end{equation}
where $p_{m}(t)$ denotes the population on chromophore $m$ at a given time $t$, $\mathbf{r}_{m}$ is the position vector of chromophore $m$ and $r(t)$ is the mean position of the excitation. 
The dynamics of this measure of spreading is shown Figure \ref{fig:spatialdistribution}. Using generalized F\"orster theory (details of which are not presented here) we have estimated that transfer times between adjacent PE545 complexes is of the order of 0.7 to 1 ps.  In this biological timescale, quantized vibrations guarantee efficient inter-complex transfer by promoting a wider distribution of excitation energy  (dotted vertical line in Figure \ref{fig:spatialdistribution}) and faster  transfer to acceptor states (see dotted vertical line Figure \ref{fig:temporaldistribution}). These conclusions also hold when a small degree of correlated fluctuations is included as shown in insets of Figures \ref{fig:temporaldistribution} and \ref{fig:spatialdistribution}. 

\subsection {Exploitation of quantum coherent dynamics to tune useful resonances} 
\begin{figure}[]
\centering 
\subfloat[Enhanced population of $\rm{PEB_{50C}}$ when coherent dynamics due to correlations is present (dashed line). Solid line indicates non-Markovian yet incoherent dynamics. In both cases only mode $\omega_{QM}=1111\rm{cm}^{-1}$ is considered]{
\includegraphics[width=85mm]{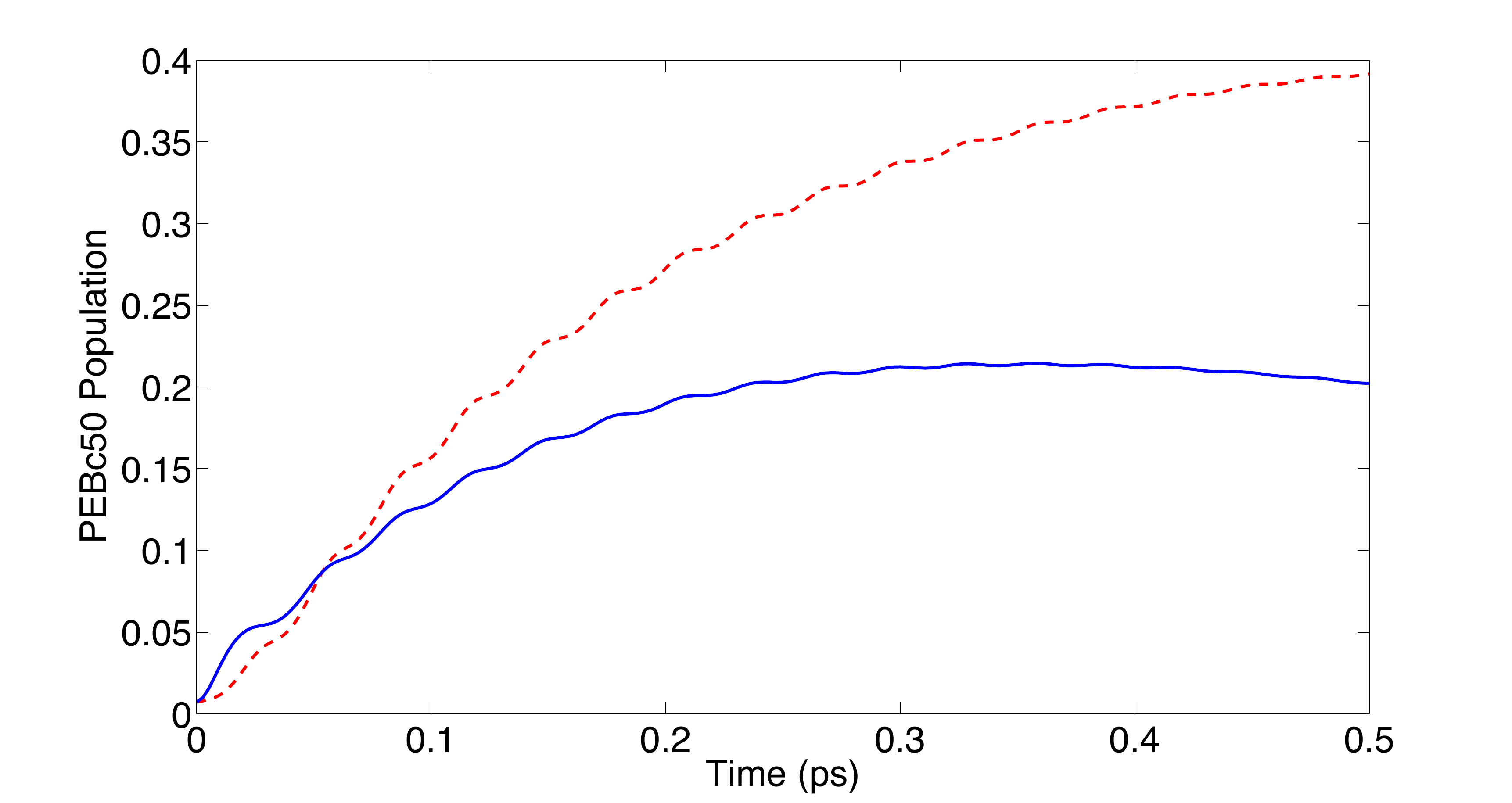}
\label{fig:activation}
}\\
\subfloat[Population of acceptor exciton state (solid line) and excitonic coherence (dashed line) for the dimer-plus-modes model]{ 
\includegraphics[width=85mm]{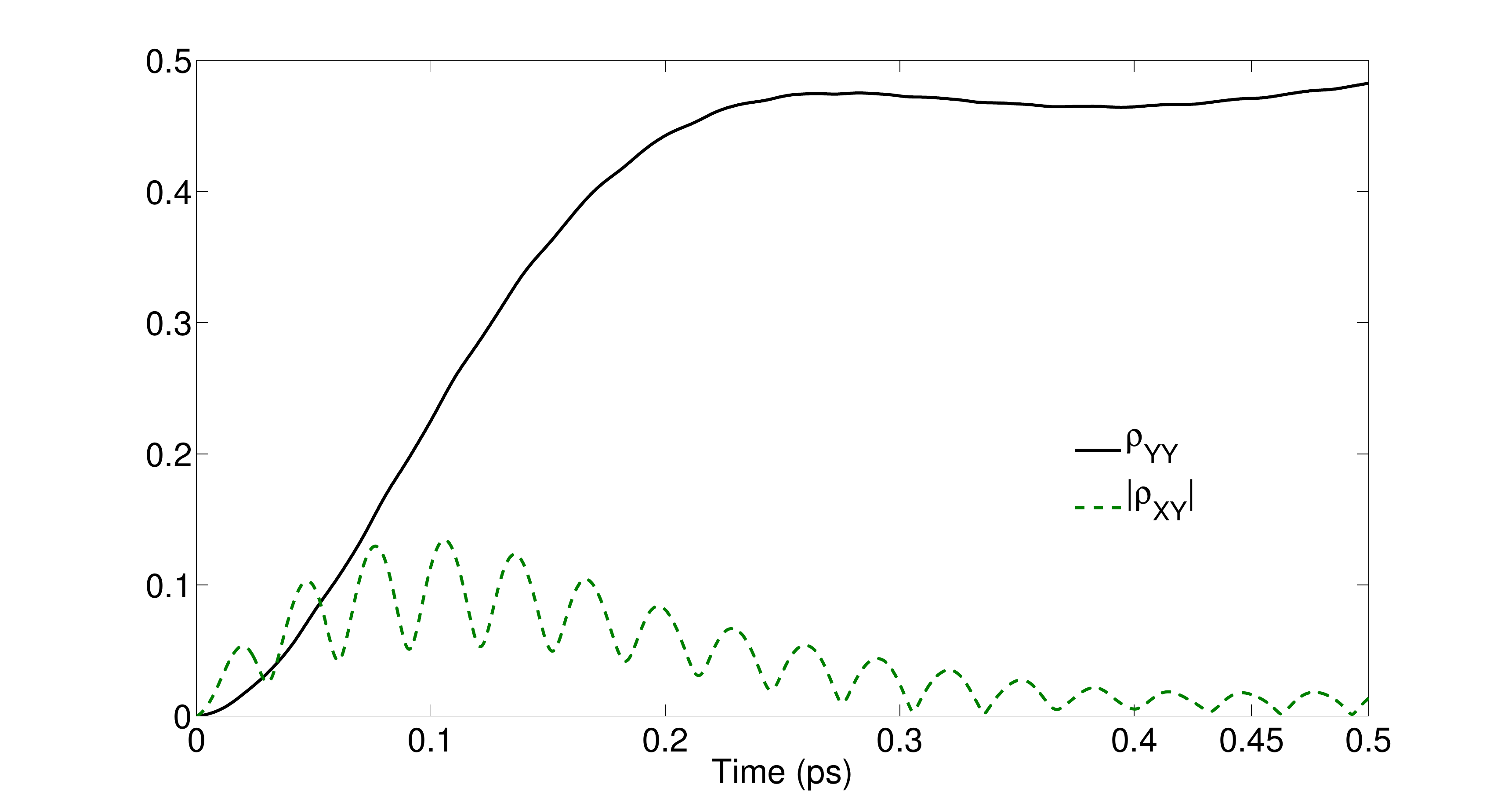}
\label{fig:toymodel}
}
\caption{Fast vibrations allow exploitation of quantum coherent evolution of excitonic coherences to enhance transport}
\end{figure}

We have discussed how a small degree of correlated fluctuations supports quantum-coherent dynamics of excitonic states and ensures tuning of a resonance with a preferred vibrational mode that enhances energy transport to a targeted site in the complex. For clarity this is summarized in Figure\ref{fig:activation} showing the enhancement of the population of the central pigment $\rm{PEB_{50C}}$ when coherent dynamics (correlations) is turned on and only the mode $\omega_{QM}=1111\rm{cm}^{-1}$ is considered. These results lead us to argue, that \textit {in the presence of quantized vibrations, any mechanism supporting coherent evolution of excitonic states, will also ensure that resonances enhancing transport are activated}. The rationale behind this idea is simple: coherent evolution effectively implies that excitonic gaps during dynamics do not fluctuate much and hence frequencies associated to coherent evolution can be synchronized with quasi-resonant vibrations for longer times, which in turn enhances population transport. 

To illustrate this idea further, we consider now the situation where quantum-coherent mixing of electronic and vibrational degrees of freedom is allowed, which generates \textit{room-temperature} coherent evolution of excitonic states. To show this we consider the full Hamiltonian evolution of an enlarged system that inlcudes both electronic states and quantized vibrations. For simplicity we consider a simplified model consisting of the central dimer formed by  $\rm{PEB}_{50C}$ and $\rm{PEB}_{50D}$ with each pigment coupled to only one mode of frequency $\omega_{QM}=1111\rm{cm}^{-1}$. As mentioned before, this mode is in close resonance with the energy difference between the two excitonic states of the dimer which we denote X and Y. To capture the effects of the external thermal background we consider each site to be subject to a small amount of pure dephasing. Details of this simplified model are presented in Appendix \ref{sec:a2}.  The initial state of the dimer-plus-modes system is separable, with the electronic part in the highest energy exciton state X (and hence no initial exciton coherence) and the high-frequency modes in thermal equilibrium. In this way, if no modes are present, there is no transport. In this situation we see that coherent exciton-mode interactions: (i) activate population transport to the low-lying excitonic state Y  and (ii) create and modulate long-lived (picosecond) excitonic coherences (Figure \ref{fig:toymodel}). Creation of excitonic coherences is, of course, expected as the initial state is not an eigenstate of the full dimer-plus-modes system. Notice that our calculations are at \textit{room-temperature} where the thermal population of the mode is mostly limited to the lowest number states, hence the importance of its quantum behavior.  The full advantages of quantum features of high-energy modes in transport processes will be soon presented in a separate publication \cite{edward}.  A recent work \cite{chin12} has also considered modulation of excitonic coherences in the Fenna-Matthews-Olson complex due a vibration whose energy is comparable to the thermal energy scale ($180~\rm{cm}^{-1}$). It would be interesting to compare the differences that can arise in exciton coherence dynamics and non-equilibrium transport properties depending of the energy scale of the vibration involved. 

\section{Concluding remarks}
The role of vibrations for adjusting and enhancing transport in wide variety of biological and chemical systems is currently of central interest \cite{west11, eisenmayer12, womick11, richards12, turner12, ritschel12, hay12}. In this article we have illustrated the importance of quantized vibrations for efficient energy transport in photosynthetic systems where excitonic states are highly localized such as the PE545  antennae protein present in cryptophyte algae. The main feature allowing these systems to take full advantage of fast vibrations is the near-resonance between vibrational frequencies and the energy gaps between excitonic states. An important implication of this close energy match is that  quantum-coherent evolutions can then tune the resonances and direct excitation energy in a preferential manner. Therefore, we argue that \textit{in photosynthetic complexes where transport is vibration-activated, a fundamental biological function for quantum coherent contributions to dynamics is to support resonances that promote fast and effective energy distribution}. 

The non-Markovian master equation in the polaron framework captures some important effects of the non-equilibrium dynamics associated to the vibrations of interest such as long-time modulation of the excitonic coherences at room temperature, predicting time scales in good agreement with those reported in experiments. However, a full quantum treatment of the interactions between excitons and these quasi-resonant  vibrations is needed both to assert the accuracy of the results here presented and to have a more complete picture of the rage of non-equilibrium effects in different chryptophyte antennae. We are currently pursuing such study and our preliminary results confirm the features presented in this article and indicate that rich nonclassical and non-equilibrium behavior arise in these systems \cite{edward}.  Atomistic calculations identifying the structural origin of such vibrations are therefore crucial. From the experimental view point it is important to develop more direct measurements of the influence of such vibrations in channelling and/or enhancing excitation energy transfer \cite{caycedo12}  as well as conclusive signatures of their quantum properties in conditions of biological relevance. The integration of physical descriptions as the ones presented in this article, together with atomistic calculations and experimental characterization will provide useful insights for the design of antennae prototypes that exploit electronic interactions and promoting vibrations for controlled and enhanced performance. 

\acknowledgements
We thank Benedetta Mennucci, Rienk van Grondelle, Elisabet Romero, Ulrich Kleinekathoefer and Carles Curuchet for discussions. A.K, E.J.O and A.O-C acknowledge funding from the EPSRC. G.D. S. thanks the Natural Sciences and Engineering Research Council of Canada, and DARPA (QuBE) for support. 

\appendix
\section{Linear Spectra Within the Polaron Formalism}
\label{sec:a1}
In this Appendix, we derive an expression for the dipole correlation function within the polaron formalism used in section \ref{sec:polaron} The dipole-dipole correlation function is defined as 

\begin{equation}
C_{d-d}(t) = \textrm{Tr}_{S+B} \{ \mu(t) \mu(0) \rho_{B} |0\rangle\langle 0| \}
\end{equation}

\noindent where $\mu(t) = e^{-i H t}~\mu~e^{i H t}$ and the system dipole operator $\mu$ is defined as the sum of the dipole operators for the individual chromophores as follows:

\begin{equation}
\mu = \sum_{m} \mu_{m} |0 \rangle\langle m| + \mu_{m}^{*} |m \rangle\langle 0|.
\end{equation}

Let us now transform into the polaron frame defined by the transformation $\tilde{H} = e^{S} H e^{-S}$, where $S=\sum_{m} \sigma_{m}^{+} \sigma_{m}^{-} \sum_{\mathbf{k}}  (h_{\mathbf{k},m} b_{\mathbf{k}}^{\dagger} - h_{\mathbf{k},m}^{*} b_{\mathbf{k}})$, with $h_{\mathbf{k},m} = g_{\mathbf{k},m}/\omega_{\mathbf{k}}$. Within the polaron frame, the dipole-dipole correlation function becomes:

\begin{equation}
C_{d-d}(t) = Tr_{S+B} \{ \tilde{\mu}(t) \tilde{\mu}(0) \tilde{\rho}_{B} |0\rangle\langle 0| \}
\end{equation}

\noindent while the dipole operator in the polaron frame is given by:

\begin{eqnarray}
\tilde{\mu} &=& \sum_{m} \Big( \prod_{\mathbf{k}} D(-h_{\mathbf{k},m}) \mu_{m} |0 \rangle\langle m| + h.c. \Big) \nonumber\\
&=& \sum_{\alpha, m} \Big( \prod_{\mathbf{k}} D(-h_{\mathbf{k},m}) \mu_{m} u_{m\alpha} |0 \rangle\langle \alpha| + h.c. \Big),
\end{eqnarray}

\noindent where $D(h_{\mathbf{k},j})=e^{h_{\mathbf{k},j} b_{\mathbf{k}}^{\dagger} - h_{\mathbf{k},j }^{*} b_{\mathbf{k}}}$ is the bath displacement operator of mode $\mathbf{k}$ due to interaction with site $j$.

Let us now proceed to calculate the dipole-dipole correlation function in steps. Firstly consider:

\begin{eqnarray}
\tilde{\mu}(0) \tilde{\rho}_{B} |0\rangle\langle 0| = \sum_{\alpha, m} \Big( \prod_{\mathbf{k}} D(-h_{\mathbf{k},m}) \mu_{m} u_{m\alpha} |0 \rangle\langle \alpha| 
+ h.c. \Big) |0\rangle\langle 0| \tilde{\rho}_{B} \nonumber\\
= \sum_{\alpha, m} \prod_{\mathbf{k}} D(h_{\mathbf{k},m}) \mu_{m}^{*} u_{m\alpha} |\alpha \rangle\langle 0| \tilde{\rho}_{B}. ~~~~~~~~~~~~~~~~~~~~~~
\end{eqnarray}

Then the dipole-dipole correlation function is:

\begin{widetext}
\begin{eqnarray}
C_{d-d}(t) = \sum_{\alpha, m} \sum_{\beta, n} \prod_{\mathbf{k}} Tr_{S+B} \{ e^{-i \tilde{H} t} \Big( D(-h_{\mathbf{k},n}) \mu_{n} u_{n\beta} |0 \rangle\langle \beta| + D(h_{\mathbf{k},n}) \mu_{n}^{*} u_{n\beta} |\beta \rangle\langle 0| \Big) e^{i \tilde{H} t} \nonumber\\ 
\times D(h_{\mathbf{k},m}) \mu_{m}^{*} u_{m\alpha} |\alpha \rangle\langle 0| \tilde{\rho}_{B} \} ~~~~~~~~~~~~~~~~~~~~~~~~~~~~~~~~~~~~~~~~~~~~~ \nonumber\\
= \sum_{\alpha, m} \sum_{\beta, n} \prod_{\mathbf{k}} Tr_{S+B} \{ \Big( D(-h_{\mathbf{k},n}) \mu_{n} u_{n\beta} |0 \rangle\langle \beta| + D(h_{\mathbf{k},n}) \mu_{n}^{*} u_{n\beta} |\beta \rangle\langle 0| \Big) ~~~~~~~~~~~~~ \nonumber\\ 
\times~ e^{i \tilde{H} t} D(h_{\mathbf{k},m}) \mu_{m}^{*} u_{m\alpha} |\alpha \rangle\langle 0| \tilde{\rho}_{B} e^{-i \tilde{H} t}\} ~~~~~~~~~~~~~~~~~~~~~~~~~~~~~~~
\nonumber\\
= \sum_{\alpha, m} \sum_{\beta, n} \prod_{\mathbf{k}} Tr_{S+B} \{ \Big( D(-h_{\mathbf{k},n}) \mu_{n} u_{n\beta} |0 \rangle\langle \beta| + D(h_{\mathbf{k},n}) \mu_{n}^{*} u_{n\beta} |\beta \rangle\langle 0| \Big) ~~~~~~~~~~~~~ \nonumber\\ 
\times~ e^{i \tilde{H} t} D(h_{\mathbf{k},m}) e^{-i \tilde{H} t} e^{i \tilde{H} t} \mu_{m}^{*} u_{m\alpha} |\alpha \rangle\langle 0| \tilde{\rho}_{B} e^{-i \tilde{H} t}\}. ~~~~~~~~~~~~~~~~~~
\end{eqnarray}
\end{widetext}

At this point, we make a number of approximations to simplify the expression for the dipole-dipole correlation function. Firstly, we shall assume that the interaction Hamiltonian $\tilde{H}_{I}$ does not perturb the bath. This amounts to assuming that the bath is in the stationary, thermal state within the polaron frame ($\tilde{\rho}_{B}$) at all times. Under this assumption we may write

\begin{eqnarray}
e^{i \tilde{H} t} D(h_{\mathbf{k},m}) e^{-i \tilde{H} t} = e^{i \tilde{H}_{0} t} D(h_{\mathbf{k},m}) e^{-i \tilde{H}_{0} t} = D(h_{\mathbf{k},m}(t)). \nonumber\\
\end{eqnarray}

The second assumption we shall make is that all system coherence terms decouple. This amounts to making the secular approximation for the system dynamics. Therefore, we may write:

\begin{equation}
e^{i \tilde{H}_{I}(t) t} \mu_{m}^{*} u_{m\alpha} |\alpha \rangle\langle 0| \tilde{\rho}_{B} e^{-i \tilde{H}_{I}(t) t} = \mu_{m}^{*} u_{m\alpha} \tilde{\rho}_{\alpha 0}(t).
\end{equation}

Under these assumptions, the dipole-dipole correlation function has the following very simple form:

\begin{eqnarray}
C_{d-d}(t) = \sum_{\alpha\beta,mn} Tr_{S} \{ \mu_{m}^{*} \mu_{n} u_{m\alpha} u_{n\beta} |0\rangle\langle \beta| \tilde{\rho}_{\alpha 0}(t) \} \nonumber\\
~~~~~~~ \times \prod_{\mathbf{k}} Tr_{B} \{ D(-h_{\mathbf{k},n}) D(h_{\mathbf{k},m}(t)) \tilde{\rho}_{B} \} \nonumber\\
= ~\sum_{\alpha} C_{\alpha}(t) \tilde{\rho}_{\alpha 0}(t). ~~~~~~~~~~~~~~~~~~~~~~~~~~~
\end{eqnarray}

\noindent where we have defined the dressed bath correlation function $C_{\alpha}(t)$ as

\begin{equation}
C_{\alpha}(t) = \sum_{mn} \mu_{m}^{*} \mu_{n} u_{m\alpha} u_{n\alpha} Tr_{B} \{ D(-h_{\mathbf{k},n}) D(h_{\mathbf{k},m}(t)) \tilde{\rho}_{B} \}.
\end{equation}

To complete the calculation of the linear spectra we need to know how each of the off-diagonal element $\tilde{\rho}_{\alpha 0}$ evolves with time. To do so, let us consider the Markovian secular polaron master equation in the interaction picture:

\begin{eqnarray}\label{eq:markov}
\frac{d\tilde{\rho}(t)}{dt}&{} ={}& - \sum_{\mu \nu} \Big( \Gamma_{\mu\nu,\nu\mu}^{(1)} \big[S_{\mu\nu}, S_{\nu\mu} \tilde{\rho}(t) \big] \:{+} \Gamma_{\mu\nu,\mu\nu}^{(2)} \big[S_{\mu\nu}^{\dagger}, S_{\mu\nu} \tilde{\rho}(t) \big] \nonumber\\ && ~~~~~~~~ \:{+} \Gamma_{\mu\nu,\mu\nu}^{(3)} \big[ S_{\mu\nu}, S_{\mu\nu}^{\dagger} \tilde{\rho}(t) \big] \:{+} \Gamma_{\mu\nu,\nu\mu}^{(4)} \big[ S_{\mu\nu}^{\dagger}, S_{\nu\mu}^{\dagger} \tilde{\rho}(t) \big]  \nonumber\\ && ~~~~~~~~~ + {\rm h.c.} \Big).
\end{eqnarray}

Details of this master equation can be found in Kolli \textit{et al.} \cite{kolli11}. The equation of motion for the off-diagonal element $\tilde{\rho}_{\alpha 0}$ is then given by:

\begin{equation}
\dot{\tilde{\rho}}_{\alpha 0}(t) = - \sum_{\mu} \Big(\Gamma_{\alpha\mu,\mu\alpha}^{(1)} + \Gamma_{\mu\alpha\mu\alpha}^{(2)} + \Gamma_{\alpha\mu,\alpha\mu}^{(3)} + \Gamma_{\mu\alpha,\alpha\mu}^{(4)} \Big) \tilde{\rho}_{\alpha 0}(t).
\end{equation}

The general solution for the evolution of the coherence terms, in the Schr\"odinger picture, is then simply given by $\tilde{\rho}_{\alpha 0}(t) = \exp(i \omega_{\alpha} t) \exp(-\tilde{\Gamma}_{\alpha} t) \tilde{\rho}_{\alpha 0}(0)$, where have defined the total dephasing rate as $\tilde{\Gamma}_{\alpha} = \sum_{\mu} \Big(\Gamma_{\alpha\mu,\mu\alpha}^{(1)} + \Gamma_{\mu\alpha\mu\alpha}^{(2)} + \Gamma_{\alpha\mu,\alpha\mu}^{(3)} + \Gamma_{\mu\alpha,\alpha\mu}^{(4)} \Big)$. Therefore, the final expression for the dipole-dipole correlation function becomes:

\begin{eqnarray}
C_{d-d}(t) = \sum_{\alpha} C_{\alpha}(t) \exp(-\tilde{\Gamma}_{\alpha} t) \exp(i \omega_{\alpha} t).
\end{eqnarray}

\section{Dimer model}
\label{sec:a2}
This model considers the central dimer in PE545 formed by pigments $\rm{PEB_{50D}}$ and $\rm{PEB_{50C}}$. 
The excitonic Hamiltonian is given by
$H_e = \epsilon_1 \sigma^+_1\sigma^-_1 + \epsilon_2
\sigma^+_2\sigma^-_2 + V(\sigma^+_1\sigma^-_2+\sigma^+_2\sigma^-_1)$ where the strength of the electronic interaction 
is $V=92~\rm{cm}^{-1}$ and the exciton eigenstates are labeled X and Y. Each pigment is coupled to a localized mode
with frequency $\omega_{QM}=1111~\rm{cm}^{-1}$. The free Hamiltonian for the modes is $H_\textrm{modes} = \omega_{QM}(b_1^\dag b_1+ b_2^\dag b_2)$
while their linear interaction with the dipole-coupled dimer is given by
$H_{e-\textrm{modes}} = g \sigma^+_1\sigma^-_1(b^\dag_1+ b_1) +
g\sigma^+_2\sigma^-_2(b^\dag_2+ b_2)$.
Here the strength of the site-mode coupling is $g=267\, \rm{cm}^{-1}$. The initial state of the full dimer plus modes system $\varrho(0) =
\rho_e(0)\otimes\rho_\textrm{modes}(0)$ is propagated under the
total Hamiltonian $H =H_e+H_\textrm{modes}+H_{e-\textrm{modes}}$.

The initial state of the dimer is chosen to be the highest energy
eigenstate X such that in the absence of the harmonic modes no transport occurs.
The modes are assumed to be initially in thermal equilibrium at room
temperature (300K). For the purposes of calculations, the Fock space
of the modes is truncated to a maximum occupation $M = 8$ which fully
captures the dynamics of the high-energy modes for the 
parameters considered. We also introduce pure dephasing of onsite energies which
results in a dissipator of the form
$
D(\varrho) = \gamma \sum_{i}^{2} (\sigma_{i}^{+}\sigma_{i}^{-} \varrho
\sigma_{i}^{+}\sigma_{i}^{-} -
\frac{1}{2}\{\sigma_{i}^{+}\sigma_{i}^{-},\varrho\})$ with $\gamma=50~\rm{cm}^{-1}$.\\

\textbf{See supplementary material for details on the polaron formalism and a table of renormalization factors of electronic couplings when solvent-induced correlations have been included}

\end{document}